\documentclass[twocolumn]{aastex631}

\usepackage{graphics,graphicx}
\hypersetup{urlcolor=blue}
\usepackage{natbib}
\usepackage[outdir=./]{epstopdf}
\usepackage{textcomp,gensymb}
\usepackage{xfrac}
\usepackage{xcolor}
\usepackage[normalem]{ulem}
\usepackage{mathrsfs}

\newcommand\kms{km~s$^{-1}$}

\newcommand{\gaia}{\textit{Gaia}}

\newcommand{\var}{$V\rm{ar}$}
\newcommand{\varg}{$V\rm{ar}_{G}$}
\newcommand{\varb}{$V\rm{ar}_{BP}$}
\newcommand{\varr}{$V\rm{ar}_{RP}$}
\newcommand{\vn}{$V\rm{ar}_{90}$}
\newcommand{\vgn}{$V\rm{ar}_{G,90}$}
\newcommand{\vbn}{$V\rm{ar}_{BP,90}$}
\newcommand{\vrn}{$V\rm{ar}_{RP,90}$}
\accepted{June 19, 2023}
\submitjournal{ApJ}

\shorttitle{Excess error and age}
\shortauthors{Barber \& Mann}
\graphicspath{{./}{}}

\begin{document}

\title{Using the Gaia excess uncertainty as a proxy for stellar variability and age
}

\correspondingauthor{Madyson G. Barber}
\email{madysonb@live.unc.edu}

\author[0000-0002-8399-472X]{Madyson G. Barber}%
\affiliation{Department of Physics and Astronomy, The University of North Carolina at Chapel Hill, Chapel Hill, NC 27599, USA}

\author[0000-0003-3654-1602]{Andrew W. Mann}%
\affiliation{Department of Physics and Astronomy, The University of North Carolina at Chapel Hill, Chapel Hill, NC 27599, USA} 

\begin{abstract}
Stars are known to be more active when they are young, resulting in a strong correlation between age and photometric variability. The amplitude variation between stars of a given age is large, but the age-variability relation becomes strong over large groups of stars. We explore this relation using the excess photometric uncertainty in \gaia\ photometry (\varg, \varb, and \varr) as a proxy for variability. The metrics follow a Skumanich-like relation, scaling as $\simeq t^{-0.4}$. By calibrating against a set of associations with known ages, we show how \var\ of population members can predict group ages within 10-20\% for associations younger than $\simeq$2.5 Gyr. In practice, age uncertainties are larger, primarily due to finite group size. The index is most useful at the youngest ages ($<$100\,Myr), where the uncertainties are comparable to or better than those derived from a color-magnitude diagram. The index is also widely available, easy to calculate, and can be used at intermediate ages where there are few or no pre- or post-main-sequence stars. We further show how \var\ can be used to find new associations and test if a group of co-moving stars is a real co-eval population. We apply our methods on the Theia groups within 350\,pc and find $\gtrsim$90\% are inconsistent with drawing stars from the field and $\simeq$80\% have variability ages consistent with those derived from the CMD. Our findings suggest the great majority of these groups contain real populations.
\end{abstract}


\section{Introduction}\label{sec:intro}

Compared to most stars, we know the age of the Sun to better than 1\% \citep{Connelly2012}. The tight age constraint comes from meteorites, rather than observations of the Sun's photosphere. Since meteorites from other stars are not available, we must rely on less precise techniques to age-date stars, such as chromospheric activity \citep[e.g.,][]{Zhou2021, Kiman2021}, rotation \citep[e.g.,][]{Barnes2007, Curtis2020}, or cooling tracks of brown dwarfs and white dwarfs \citep[e.g.,][]{Kilic2019, Marley2021}.

Outside the Sun, stars with the most precise and reliable ages are usually in co-eval associations \citep{Soderblom2014}. Ages can then be estimated using the bulk properties of the cluster, such as the lithium abundances \citep[e.g.,][]{Burke2004, Wood2022} or main-sequence turn-off \citep{Conroy2010}, or from a subset of stars with more easily determined properties \citep[e.g., asteroseismic pulsators;][]{Grunblatt2021, Bedding2022}. 

Precision astrometry from the \gaia\ mission \citep{Gaia_mission2016} has been invaluable for finding new stellar associations \citep[e.g.,][]{Meingast2019, Moranta2022}, sub-populations of known associations \citep[e.g.,][]{Wood2022}, and additional members of known populations \citep[e.g.,][]{Gagne2018,Roser2020}. Identifying and finding members of sparse groups is still challenging. Galactic shear causes the group's velocity dispersion to grow with time \citep{Dobbs2013}. Larson's laws also imply that groups with a larger spatial scale should also exhibit a larger velocity spread \citep{Larson1981}, and the resulting velocity dispersion can exceed typical measurement uncertainties from \gaia. Further, the more the population extends spatially, the greater the number of nearby field stars that will align with the group's kinematics by chance. 

To aid with search and selection, many studies add an additional requirement to select on, such as a color-magnitude (CMD) position consistent with being pre-main-sequence \citep[e.g.][]{Kerr2021} or spectroscopic indicators of activity \citep[e.g.,][]{Zerjal2021}. These are often observationally expensive and/or only apply to a subset of stars. Thus, additional metrics would be invaluable when searching for young stellar associations. An activity metric that is already widely available would be particularly useful for mining all-sky surveys for young associations. 

\citet{Guidry2021} show that excess uncertainties in \gaia\ photometry is an indicator of source variability. They use a metric for excess uncertainty ($V_G$) to identify white dwarfs on the ZZ Ceti instability strip. \citet{Barlow2022} use the same method to identify highly variable hot subdwarfs, and \citet{Wilson2023} used a similar Gaia EDR3 G-band variability as a parameter for identifying Class II YSOs. This method for finding variable stars is not unique to DR3, and similar approaches have been used with DR1 and DR2 \citep[e.g.,][]{Belokurov2017, Vioque2020}.

The metric could be expanded to identify young stars out to the limits of \gaia. \gaia\ photometry can achieve a precision of 30\,mmag per epoch and 2\,mmag total, ($G=19$) with a typical target getting observations every few weeks \citep{Hodgkin2021}, more than sufficient to detect stellar variations expected from $<$1\,Gyr stars \citep{Rizzuto2017, Miyakawa2021}. 

Starspot coverage is known to follow a Skumanich-like decrease with age \citep{Morris2020}. The relation between starspot coverage and (observed) stellar variability is complex due to both variations in stellar inclination and astrophysical variation between stars. However, the two should be strongly correlated over large collections of stars \citep{Luger2021}. In the youngest stars, stellar variability may be driven by effects other than starspots, such as accretion \citep{Park2021} and dippers \citep{Cody2014, Ansdell2016a}, but the overall variability is still expected to be stronger with decreasing age. Thus, $V_G$ or a similar variability diagnostic could be used to provide age estimates for populations of stars.  

In this paper, we update the variability metric put forth in \citet{Guidry2021}, including extending its use to all three \gaia\ filters (Section~\ref{sec:obs}). Using a set of stars in associations with well-determined ages (Section~\ref{sec:target}), we provide a relation between the distribution of \var\ for stars in a co-eval group and the age of the group (Section~\ref{sec:calibration}). We discuss the impact of additional effects, like the distance to the population and field star contamination, in Section~\ref{sec:tests}. To highlight the power of \var, we show how it can be used to assign ages to newly identified populations of stars, test if a candidate group of co-moving stars represents a real young population, and find new associations (Section~\ref{sec:application}). 

\section{Gaia excess variability}\label{sec:obs}

\gaia\ mean flux (\texttt{PHOT\_G\_MEAN\_FLUX} or $<G>$) and uncertainty (\texttt{PHOT\_G\_MEAN\_FLUX\_ERROR} or $\sigma_{<G>}$) is calculated using the uncertainty on the weighted mean of included observations \citep{Evans2018, Riello2021}. For a non-variable source and fixed instrumental noise, $\sigma_{<G>}^2$ scales with the source flux and inversely with the number of observations ($n_{obs,G}$). Thus, a deviation above this scaling is a sign of astrophysical variation in the flux. \citet{Guidry2021} take advantage of this to identify variable white dwarfs in \gaia\ photometry, using a variability metric defined as:
\begin{equation}
    V_G \equiv \frac{\sigma_{<G>}}{<G>}\sqrt{n_{obs,G}}.
\end{equation}
A higher $V_G$ would indicate a source with more flux variation than expected from noise alone. In practice, instrumental noise varies with source brightness. \citet{Guidry2021} handled this by subtracting out the baseline relation between $V_G$ and $G$. 

Our approach to removing the scaling with brightness was to use the fitted \gaia\ photometric uncertainties tool\footnote{\url{https://github.com/gaia-dpci/gaia-dr3-photometric-uncertainties}} \citep{Riello2021_phot}. These relations were derived empirically, and hence included a wide range of effects. The code provides a predicted magnitude uncertainty ($\sigma_{G,p}$) as a function of \gaia\ $G$ magnitude and $n_{obs,G}$. We defined a new variability index we call \varg:
\begin{equation}\label{eqn:varg}
   Var_G =  \log_{10}\left({\frac{2.5}{\ln{10}}}{\frac{\sigma_{<G>}}{<G>}}\right) - \log_{10}[  {\sigma_{G,p}(G,n_{obs,G})}],
\end{equation}
The numerical factor in the first term ($\frac{2.5}{\ln{10}<G>}$) serves to convert the \gaia\ flux uncertainty ($\sigma_{<G>}$) provided in the \gaia\ catalog into magnitude space, then matching the units of the second term (the output of the fitted uncertainties code). One could rewrite $Var_G$ into a single term inside the logarithm and convert the fitted uncertainties output into a flux. In this case, the term inside the logarithm would be the ratio of the reported (flux) uncertainties to that expected for a typical source.

We extended Equation~\ref{eqn:varg} to the other two \gaia\ photometric bands, yielding \varb\ and \varr\ with a simple substitution. 

\begin{figure}[tb]
    \centering
    \includegraphics[width=0.49\textwidth]{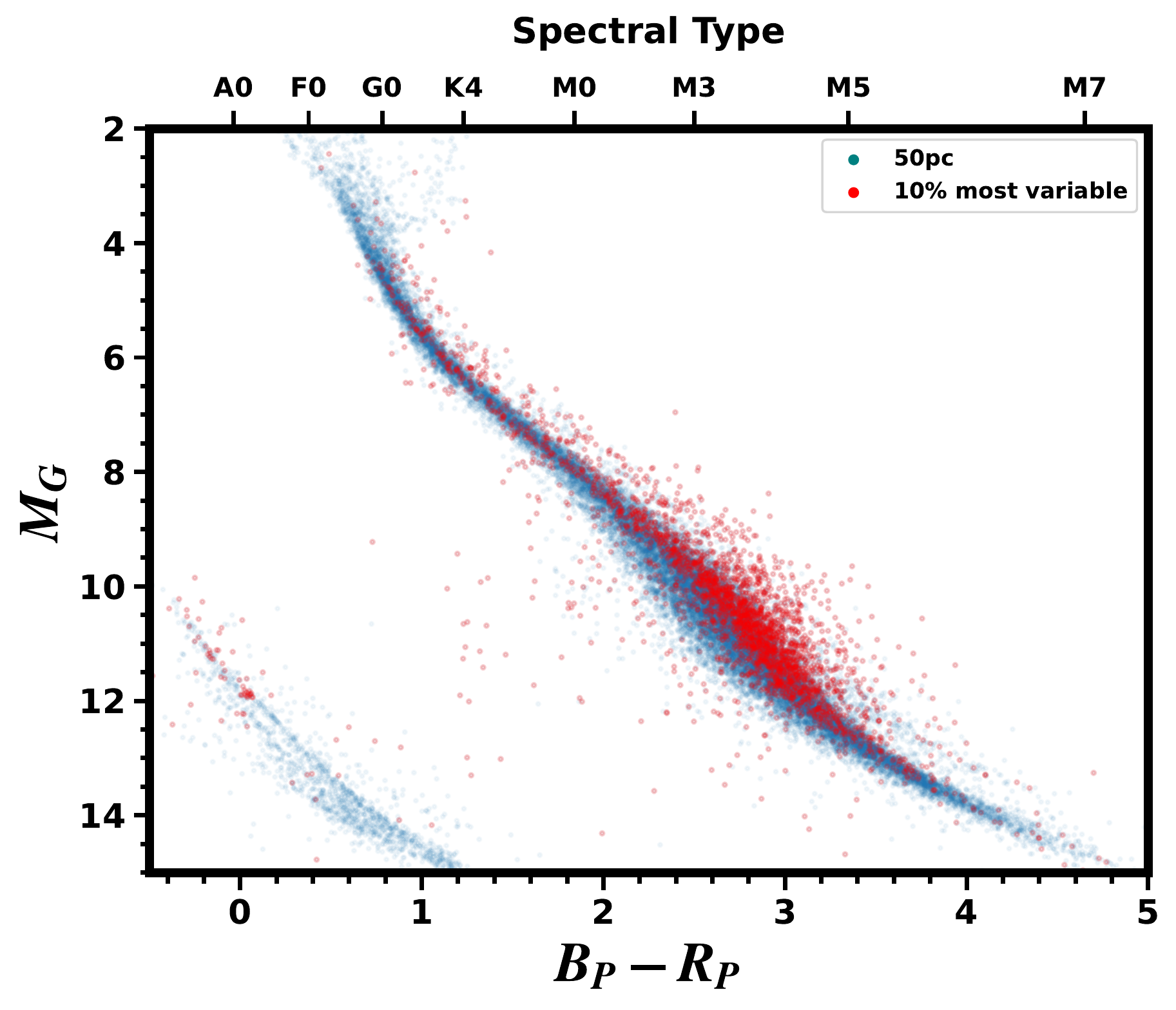}
    \caption{A color-magnitude diagram of stars within 50\,pc of the Sun (teal). Red points indicate those with \varg\ values in the top 10\% of the sample. These stars are preferentially high on the CMD for the early M dwarfs and along the zero-age main-sequence for the GK dwarfs, where we expect to see young stars.}
    \label{fig:cmd}
\end{figure}

We performed a quick demonstration that the revised metric works for young stars by checking the distribution of \varg\ with their position on the color-magnitude diagram (CMD), which we show in Figure~\ref{fig:cmd}. Stars that have the highest 10\% of \varg\ values are highlighted. As expected, these variable stars land preferentially in regions of the CMD where we see younger stars (e.g., pre-main-sequence regions for early-to-mid M dwarfs). 

\section{Target Selection}\label{sec:target}

Our goal was to find a set of groups for calibrating the relation between \var{} and age. To this end, we selected a set of co-eval populations (e.g., open clusters, moving groups, and star-formation regions) with well-determined ages and membership lists in the literature. As a comparison set and to test the effects of contamination, we also used a volume-limited sample of stars in the Solar neighborhood (random ages). We then selected the subset of stars in these groups or the field sample where \var\ is most effective. 

\subsection{Young Associations}

We restricted our calibration sample of young associations to groups within 350\,pc of the Sun. As discussed in Section~\ref{sec:calibration}, the \var\ index is distance dependent. We also found that groups past 350\,pc tended to have smaller membership lists, more uncertain ages, and more discrepant ages between literature sources. 

We required groups to have at least 40 stars after all cuts on the membership list (described in Section~\ref{sec:star}). The method works for smaller samples of stars, but the larger uncertainties makes such groups ineffective for calibration. 

The majority of our sample was taken from the sample of open clusters in \citet{Cantat-Gaudin2018} and \citet{Cantat-Gaudin2020_ages}. We added in several well-characterized clusters like 32 Ori \citep{Luhman2022}, as well as more diffuse groups like Psc-Eri\footnote{Meingast-1} \citep{Meingast2019} and $\mu$ Tau \citep{Gagne2000}.

To sample the youngest ages, we added in young associations Taurus-Auriga \citep{Krolikowski2021}, the three major groups in the Scorpius-Centaurus OB association \citep[Upper Scorpius, Upper Centaurus–Lupus, and Lower Centaurus–Crux, ][]{Preibisch2008}, the Chamaeleon complex \citep[Cha I and Cha II;][]{Luhman2007}, and Corona-Australis \citep{Galli2020}. Earlier studies have shown that these associations are not single-aged populations. For example, \citet{Goldman2018} demonstrated that Lower Centaurus–Crux is comprised of at least four sub-populations with ages that differ by 1-3\,Myr. However, this spread is comparable to or smaller than our assigned age uncertainties. The spread between sub-population ages was only a problem for Taurus-Aurgia, where we opted to only include the youngest ($<$10\,Myr) subgroups from \citet{Krolikowski2021}. 

In total, we used 32 groups ranging in age from 3\,Myr to 2.7\,Gyr. Only one group was older than 1\,Gyr (Ruprecht 147) and more than half the groups were less than 100\,Myr. We list all selected associations in Table~\ref{tab:sample}. 

\subsubsection{Excluded groups}
Our list of associations was meant to be representative of groups near the Sun, not complete. The most common reason to skip a group was that it did not satisfy the 40-member minimum. Since we only included stars with $B_P-R_P<2.5$ (see Section~\ref{sec:star}), the full population size needed to be significantly larger. This minimum removed many young moving groups like Columba and low-mass clusters like Ursa Major and Platais 10. 

\begin{figure*}[tb]
    \centering
    \includegraphics[width=.99\textwidth]{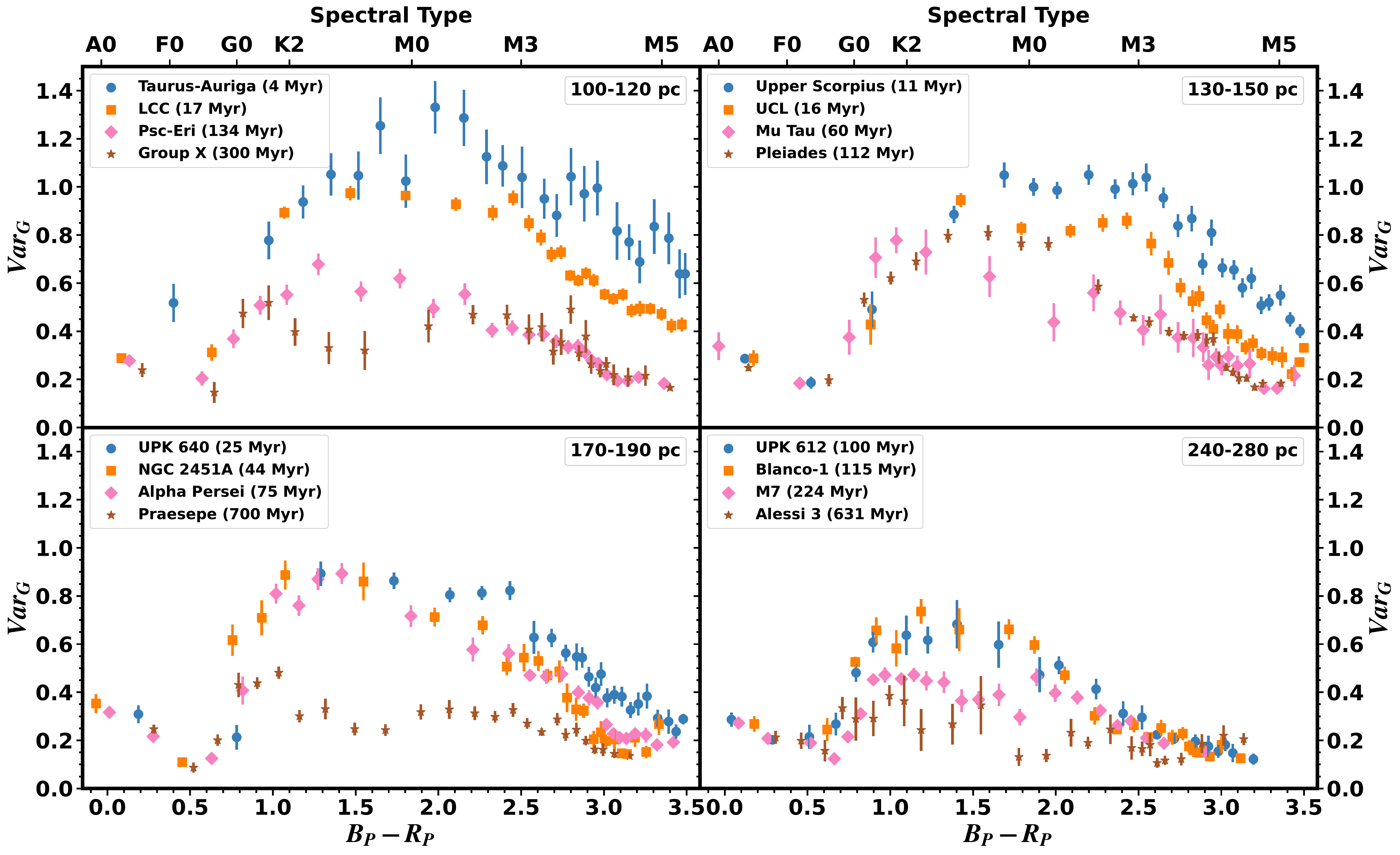}
    \caption{Running median of \varg{} as a function of color separated into similar distances: 100-120pc (top-left), 130-150pc (top-right), 170-190pc (bottom-left) and 240-280pc (bottom-right). Displayed uncertainties are the standard error on the median. Bin sizes have an equal number of stars within a group but not between groups. These show the expected sequence in age for FGK and early M dwarfs, i.e., the youngest groups have the highest \varg\ values. But for mid-to-late M dwarfs, the correlation breaks down for groups older than $\simeq$50 Myr, independent of distance.}
    \label{fig:colorseq}
\end{figure*}

Some groups were excluded because of ambiguity in the assigned age or membership. For example, Alessi 13 ({\ensuremath{\chi}}$^{1}$ For) has been assigned ages ranging from 30\,Myr \citep{Galli2021_chifor} to more than 500\,Myr \citep[e.g.][]{Yen2018}. This also led us to exclude some nearby moving groups (e.g., AB Dor, Carina-Near, and Argus), many of which have discrepant ages and membership lists in the literature \citep[e.g.,][]{Mamajek2016}. 

Newly identified groups from SPYGLASS \citep{Kerr2021} have a sample selection that is problematic for our purposes. The initial selection included only pre-MS stars, so it was heavily biased towards late-type stars where \var\ is less effective (Figure~\ref{fig:colorseq}). Their final selection had more FGK stars but suffered from higher contamination. SPYGLASS groups were also restricted to those $<$50\,Myr, where we already had 14 groups in our calibration set.

We did not include MELANGE \citep{THYMEV} and Theia \citep{Kounkel2020} groups in our calibration set. The Theia groups contain real co-eval populations \citep{Andrews2022}, but many remain controversial \citep{Zucker2022b}. Instead, we used the techniques discussed in this paper to test the existence and ages that were assigned to these sets of groups in Section~\ref{sec:application}. 

\begin{deluxetable*}{lclclc}
\centering
\tabletypesize{\scriptsize}
\tablewidth{0pt}
\tablecaption{Young Associations for Calibration \label{tab:sample}}
\tablehead{\colhead{Name} & \colhead{Age} & \colhead{Age} & \colhead{N$_{\rm{stars}}$\tablenotemark{a}} & \colhead{Membership} & \colhead{Distance\tablenotemark{b}}\\
\colhead{} & \colhead{(Myr)} & \colhead{Reference} & \colhead{} & \colhead{Reference} & \colhead{pc}
}
\startdata
Taurus-Auriga & $3.5\pm2.5$ & \citet{Krolikowski2021} & 137 & \citet{Krolikowski2021} & 145\\
Chamaeleon & $4\pm2$ & \citet{Luhman2007} & 45 & \citet{Galli2021_cha} & 191\\
Corona-Australis  & $6\pm4$ & \citet{Galli2020} & 88 & \citet{Esplin2022} & 151\\
Upper Scorpius & $11\pm3$ & \citet{Pecaut2012} & 377 & \citet{Luhman2020} & 144\\
Upper Centaurus–Lupus & $16\pm1$ & \citet{Pecaut2012} & 169 & \citet{Damiani2019} & 175\\
Lower Centarus Crux & $17\pm1$ & \citet{Pecaut2012} & 459 & \citet{Goldman2018} & 113\\
UPK 422 & $19\pm2$ & \citet{Cantat-Gaudin2020_ages} & 40 & \citet{Cantat-Gaudin2020_ages} & 300\\
32 Ori & $21\pm4$ & \citet{Luhman2022} & 46 & \citet{Luhman2022} & 103\\
UPK 640 & $25\pm3$ & \citet{Cantat-Gaudin2020_ages} & 145 & \citet{Cantat-Gaudin2020_ages} & 176\\
Platais 8 & $30\pm3$ & \citet{Cantat-Gaudin2020_ages} & 61 & \citet{Cantat-Gaudin2018} & 135\\
NGC 2232 & $38\pm3$ & \citet{Binks2021} & 94 & \citet{Cantat-Gaudin2018} & 321\\
NGC 2451A & $44\pm2$ & \citet{Bossini2019} & 121 & \citet{Cantat-Gaudin2018} & 192\\
Collinder 135 & $45\pm5$ & \citet{Kovaleva2020} & 164 & \citet{Cantat-Gaudin2018} & 299\\
IC 2602 & $46^{+6}_{-5}$ & \citet{Dobbie2010} & 99 & \citet{Cantat-Gaudin2018} & 151\\
Platais 9 & $50\pm5$ & \citet{Cantat-Gaudin2020_ages} & 51 & \citet{Cantat-Gaudin2018} & 184\\
IC 2391 & $51^{+5}_{-4}$ & \citet{Nisak2022} & 78 & \citet{Cantat-Gaudin2018} & 151\\
$\mu$ Tau & $60\pm7$ & \citet{Gagne2000} & 122 & \citet{Gagne2020} & 155\\
$\alpha$ Persei & $75^{+6}_{-7}$ & \citet{Galindo-Guil2022} & 318 & \citet{Cantat-Gaudin2018} & 174\\
UPK 612 & $100\pm10$ & \citet{Cantat-Gaudin2020_ages} & 141 & \citet{Cantat-Gaudin2020_ages} & 229\\
Pleiades & $112\pm5$ & \citet{Dahm2015} & 391 & \citet{Cantat-Gaudin2018} & 136\\
Blanco-1 & $115\pm10$ & \citet{GaiaCollaboration_2018_HR} & 195 & \citet{Cantat-Gaudin2018} & 237\\
Psc-Eri/Meingast-1 & $134\pm7$ & \citet{Roser2020} & 581 & \citet{Ratzenbock2020} & 131\\
Platais 3 & $208^{+122}_{-42}$ & \citet{Bossini2019} & 54 & \citet{Cantat-Gaudin2018} & 178\\
M7 & $224\pm22$ & \citet{Cantat-Gaudin2020_ages} & 771 & \citet{Cantat-Gaudin2018} & 280\\
Alessi 9 & $282^{+28}_{-29}$ & \citet{Cantat-Gaudin2020_ages} & 118 & \citet{Cantat-Gaudin2018} & 209\\
Group X & $300\pm50$ & \citet{THYMEVII} & 132 & \citet{Tang2019, THYMEVII} & 104\\
NGC 7092 & $310^{+74}_{-58}$ & \citet{Bossini2019} & 125 & \citet{Cantat-Gaudin2018} & 297\\
Alessi 3 & $631\pm63$ & \citet{Cantat-Gaudin2020_ages} & 171 & \citet{Cantat-Gaudin2018} & 279\\
Hyades & $650\pm70$ & \citet{Martin2018} & 283 & \citet{Roser2019, Jerabkova2021} & 134\\
Praesepe & $700\pm25$ & \citet{Cummings2018} & 422 & \citet{Cantat-Gaudin2018} & 185\\
Coma Ber & $750^{+50}_{-100}$ & \citet{Tang2018, Singh2021} & 98 & \citet{Tang2019} & 86\\
Ruprecht 147 & $2670^{+390}_{-550}$ & \citet{Torres2020} & 156 & \citet{Cantat-Gaudin2018} & 306\\
\enddata
\tablenotetext{a}{$N_{\rm{stars}}$ denotes the number of stars used in our analysis (after applying all cuts). The full membership list is always larger.}
\tablenotetext{b}{Median distance of included members.}
\end{deluxetable*}

\subsubsection{Assigning ages}\label{sec:ages}

Most of the groups used in our analysis had multiple age determinations in the literature. In order of priority, we adopted ages based on 1) the lithium depletion boundary, 2) an isochrone/CMD fit using eclipsing binaries or other benchmark stars, 3) an isochrone/CMD fit using \gaia\ data, 4) an isochrone/CMD fit using other datasets. We excluded references where no uncertainty was provided. When multiple sources with the same ranking above provided an age, we used the more precise analysis. The only deviation from this procedure was for Praesepe, for which \citet{Bossini2019} reported an unrealistic age uncertainty of only 3-4\,Myr (better than 1\%). Instead, we adopted the age from \citet{Cummings2018}. The reference used for each association age is listed in Table~\ref{tab:sample}.

\citet{Cantat-Gaudin2020_ages} derived ages using an artificial neural network run on the CMD from \gaia\ data. Using a validation set of clusters, they estimated uncertainties were 10-20\%, depending on the group size. We adopted the low end (10\% uncertainties), as most groups considered here had sufficiently large membership lists.

\subsection{Field Sample}\label{sec:field}

As a comparison set and to test how field contamination impacts \var\ in a group, we used a sample of nearby field stars from the \gaia\ catalog of nearby stars \citep{GaiaCollaboration2021}. We pulled stars from the `selected objects' within 50\,pc ($\pi>20$\,mas). 

\subsection{Star Selection}\label{sec:star}

We drew our sample of stars from the membership lists listed in Table~\ref{tab:sample} with the following cuts:
\begin{itemize}
    \item \texttt{phot\_g\_mean\_flux\_over\_error}$>30$
    \item \texttt{phot\_bp\_mean\_flux\_over\_error}$>20$
    \item \texttt{phot\_rp\_mean\_flux\_over\_error}$>20$
    \item \texttt{parallax\_over\_error}$>20$
    \item Membership probability (if provided) $>50\%$
    \item $M_G<10$ or $B_P-R_P>1$
    \item $B_P-R_P<2.5$ 
\end{itemize}
The first five restrictions removed sources with unreliable photometry or membership. Many membership lists also used quality cuts similar to the first four, so this kept the stellar sample more homogeneous between groups. Field contamination has a weak impact on our findings (see Section~\ref{sec:tests}). However, many lists contain sources with membership probability down to $\simeq0\%$, so a minimum cut was required. The sixth requirement removed any white dwarfs from the sample. 

As we show in Figure \ref{fig:colorseq}, \var\ becomes ineffective for mid-to-late M dwarfs older than $\simeq50$\,Myr. At the youngest ages, stars with $B_P-R_P>2.5$ and cooler follow the expected sequence; young groups have higher \varg. In older groups, these stars all have similar \varg\ levels independent of distance. This holds even if we consider just the nearest groups, suggesting the effect is not purely due to differences in brightness. Weakening sensitivity of \varg\ was the major motivation for the color requirement.

\section{Calibration}\label{sec:calibration}

\begin{figure}
    \centering
    \includegraphics[width=0.49\textwidth]{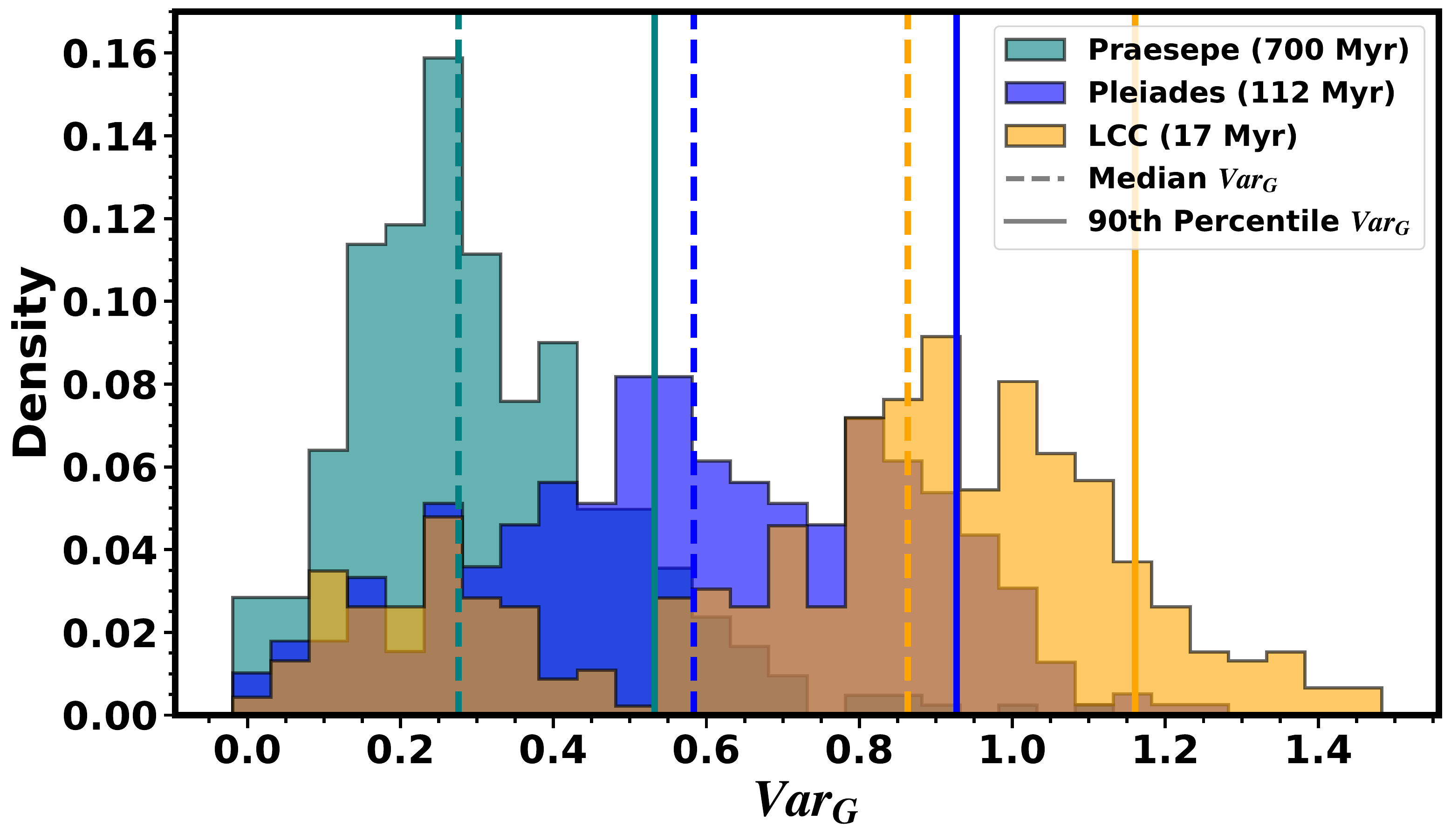}
    \caption{The distribution of \varg\ of members of Praesepe, Pleiades, and Lower Centarus Crux (LCC). For each group, solid and dashed lines indicate the median and 90th percentile, respectively. While the overall \varg\ distribution for a group clearly depends on age, there is significant overlap. Even the youngest groups have a significant population of low ($<0.4$) \varg\ stars.}
    \label{fig:groupsHistograms}
\end{figure}

As we show in Figure~\ref{fig:groupsHistograms}, \varg\ exhibits a large variation across stars in a co-eval association. As a result, there is a significant overlap in the \varg\ values between associations of different ages. If a star's \varg\ is high, it is likely to be young, but even the youngest groups show some stars with low \varg. As a result, the metric is not as useful for assigning ages to individual stars. Fortunately, the \varg\ distributions are well-sorted according to age, meaning we can make use a population-level metric to estimate the age of a group.

\begin{figure}[tb]
    \centering
    \includegraphics[width=0.49\textwidth]{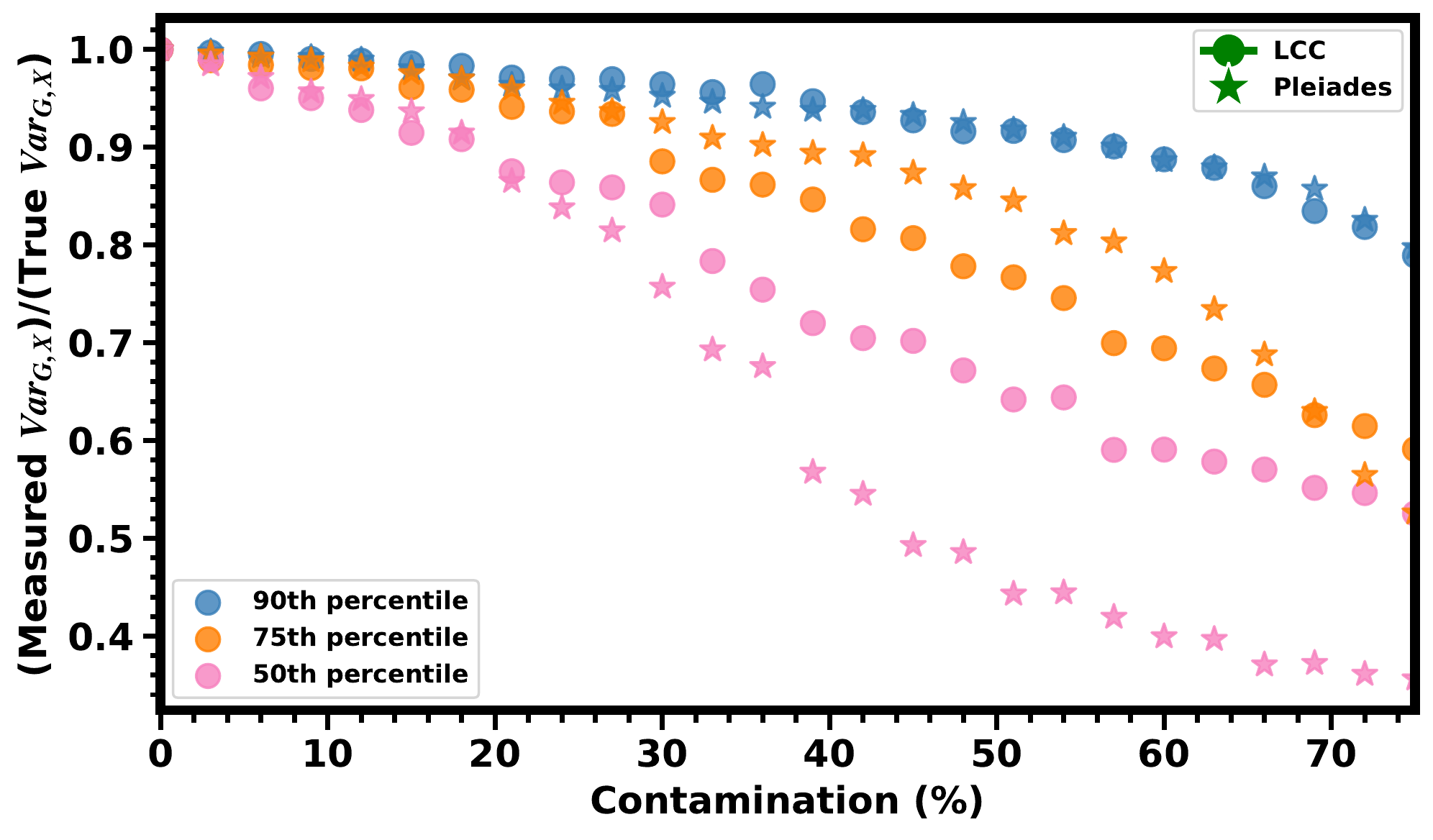}
    \caption{The $Var_{G,X}$ we would measure if a fraction of stars are field interlopers (contaminants), normalized to the value assuming no contamination. Three different values for $X$ (50\%, 75\% and 90\%) are shown as three colors. This was built by using member lists from the Pleiades (stars) and Lower Centaurus–Crux (circles), adding in nearby non-members and recomputing $Var_{G,X}$. This assumes the original list has low contamination. For both groups, contamination has a weak effect ($<20$\%) on \vgn.  }
    \label{fig:contam}
\end{figure}

For our metric, we used the 90$^{th}$ percentile (highest) \var\ value within an association. We also tested using the 50$^{th}$ (the median) and 75$^{th}$ percentile, both of which showed a strong correlation with age. We opted for the 90$^{th}$ because it showed the lowest scatter around a linear fit and exhibited a high resiliency to field-star contamination (see Figure~\ref{fig:contam} and discussion in Section~\ref{sec:tests}). We denote this value as \vn (\vgn, \vbn, and \vrn) to separate from \var, which is the metric for a single star.

We estimated uncertainties on \vn\ for each group based on a bootstrap re-sampling of the association members. For this, we used scipy's \texttt{bootstrap} with the default settings. We assumed symmetric uncertainties for simplicity.

We fit the relation between age and variability in log-log space-based on previous work relating variability to age \citep[e.g.,][]{Morris2020, Luger2021}. The \var\ parameter is equivalent to a magnitude and hence was already a log of the flux variation. Age uncertainties roughly scaled with age, and we found the fit uncertainties were better modeled as a fractional error than an absolute error (favoring working in log space). This yielded a linear relation:
\begin{equation} \label{eqn:fit}
    \log_{10}({\rm{age}})\ \mathrm{[Myr]} = m\times Var_{90} + b,
\end{equation}
where $m$ and $b$ were fit parameters. We fit this three times, once for each of the \gaia\ bandpasses (\varg, \varb, and \varr). Adding a second-order term in \vn\ gave negligible improvement on the fit, but we explored adding a distance term (Section~\ref{sec:tests}). 

We included a third fit parameter, $\ln{f}$, to capture the intrinsic scatter in the relation. This could also be interpreted as underestimated uncertainties in the input ages, but as we show in Section~\ref{sec:tests}, the result was robust to changes in the input age uncertainties. In addition, this parameter acted as a lower limit on the age uncertainties achievable with the method.

Because there are uncertainties in both \var\ and age, we use a likelihood that propagates the uncertainties in \vn\ into age uncertainties, and includes an extra term to account for the intrinsic scatter in age. This method (including the likelihood) is described in \citet{Tremaine2002} and Eqn~24 of \citet{Kelly2007}. \citet{Kelly2007} also describe some drawbacks of this method, but tests of \citet{Kelly2007}'s preferred method (\texttt{linmix}) yielded nearly identical results.

We used a likelihood maximization in a Monte Carlo Markov Chain (MCMC) schematic with \textbf{emcee} \citep{Foreman-Mackey2013}, optimizing on $\ln{\mathscr{L}}$. For each of the three filters, we adopted uniform priors on all parameters with large bounds to prevent runaway walkers ($f>0$ and $-4<m<-1$). We initialized the three parameters based on the results of least-squared fits for each filter. We then ran the chain using 30 walkers until it passed 50 times the autocorrelation time \citep[usually sufficient for convergence, ][]{goodman2010}, typically $\simeq5,000$ steps. For the burn-in, we used 10\% of the total number of steps, although the result was not sensitive to the choice of burn-in. 

\begin{figure}[tb]
    \centering 
    \includegraphics[width=0.49\textwidth]{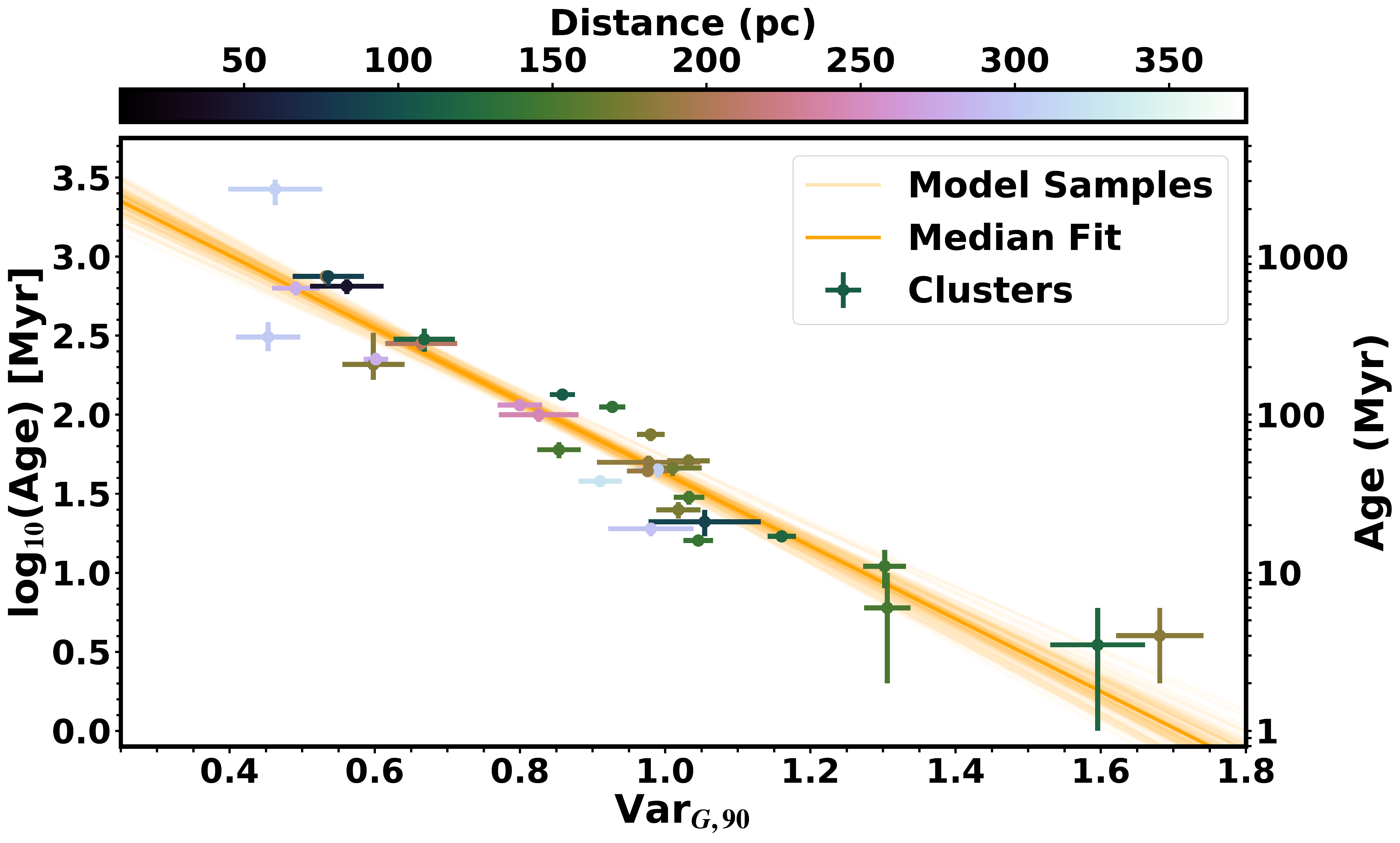}\\
    \includegraphics[width=0.49\textwidth]{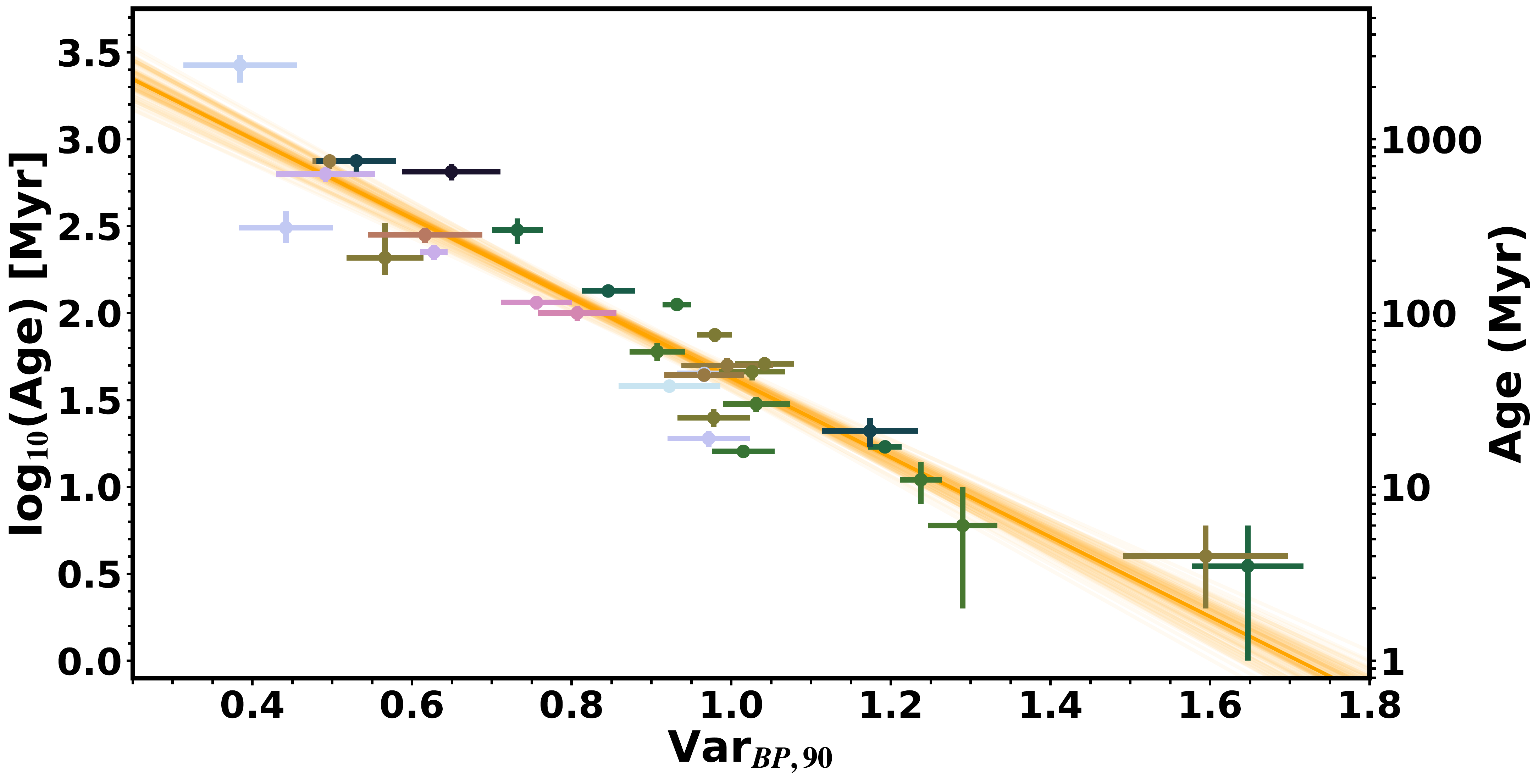}\\
    \includegraphics[width=0.49\textwidth]{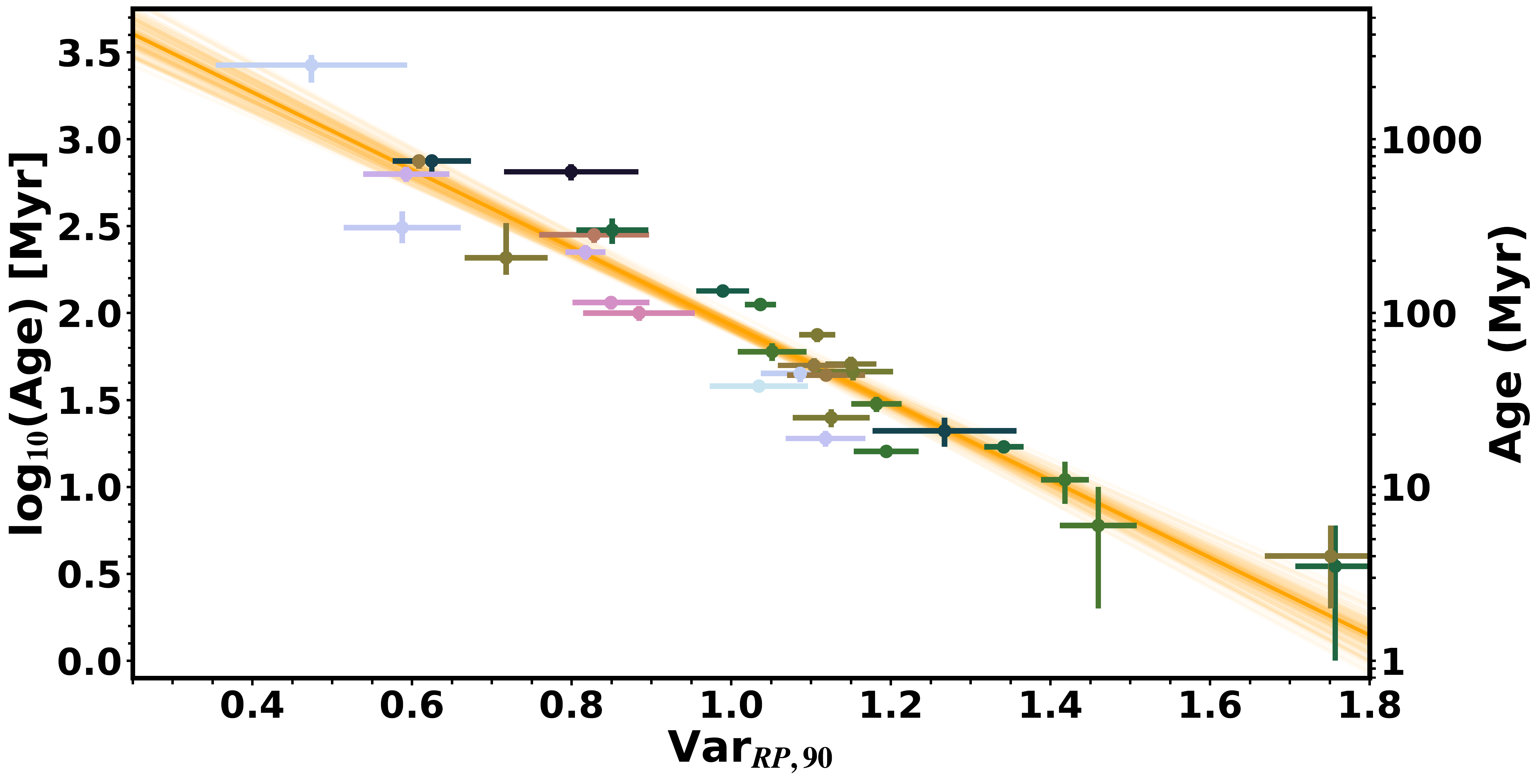}
    \caption{Age of associations as a function of Var$_{G}$ (top), Var$_{Bp}$ (middle), and Var$_{Rp}$ (bottom) using the young associations listed in Table \ref{tab:sample}. Each association is colored by its distance from the Sun. The orange line represents the best-fit for each filter, with 100 translucent orange lines of randomly drawn sample fits from the MCMC posterior. The best-fit parameters parameters are listed in Table \ref{tab:fitParams}.}
    \label{fig:caliFits}
\end{figure}

\begin{figure*}[tb]
    \centering
    \includegraphics[width=0.32\textwidth]{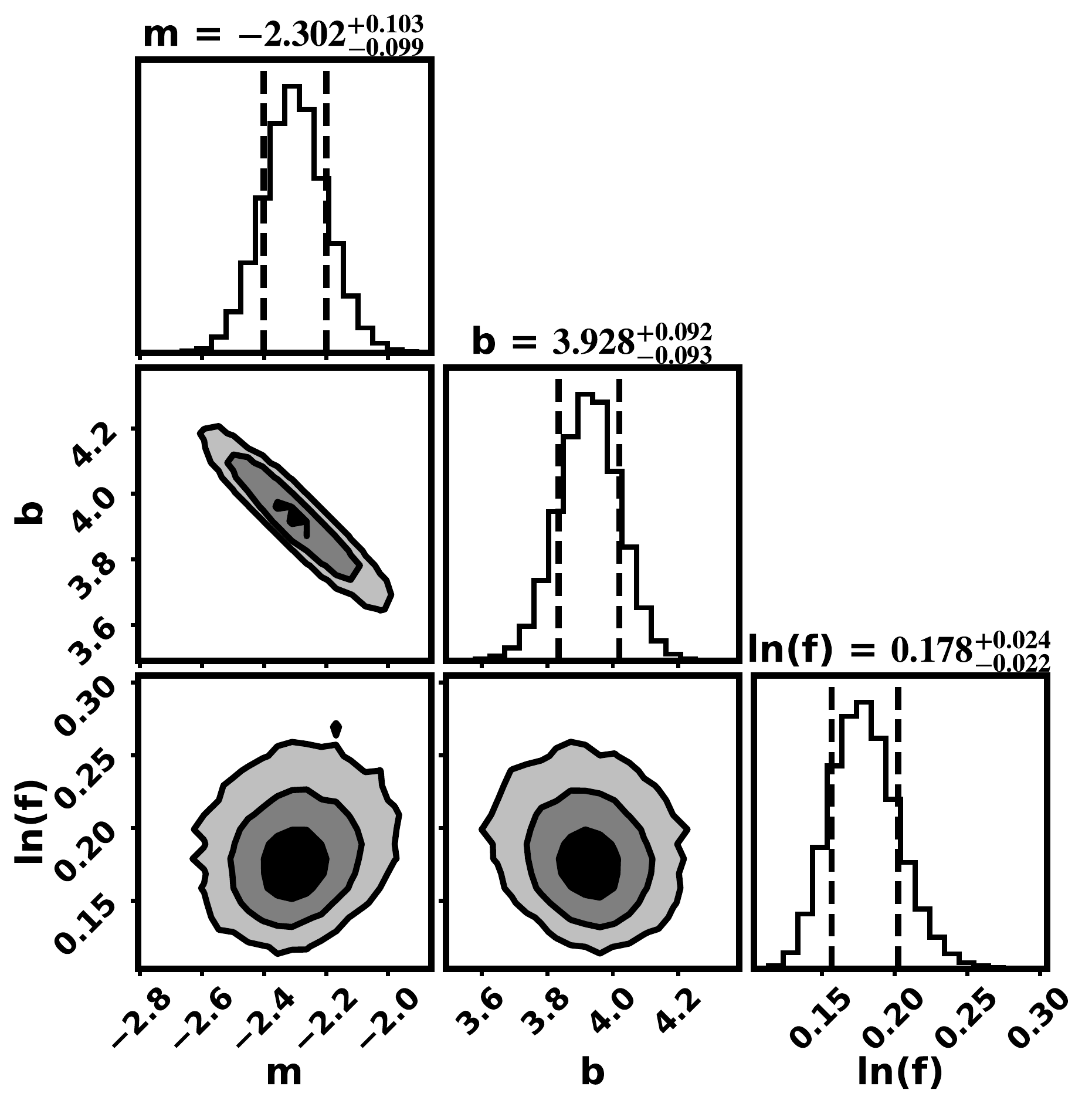}
    \includegraphics[width=0.32\textwidth]{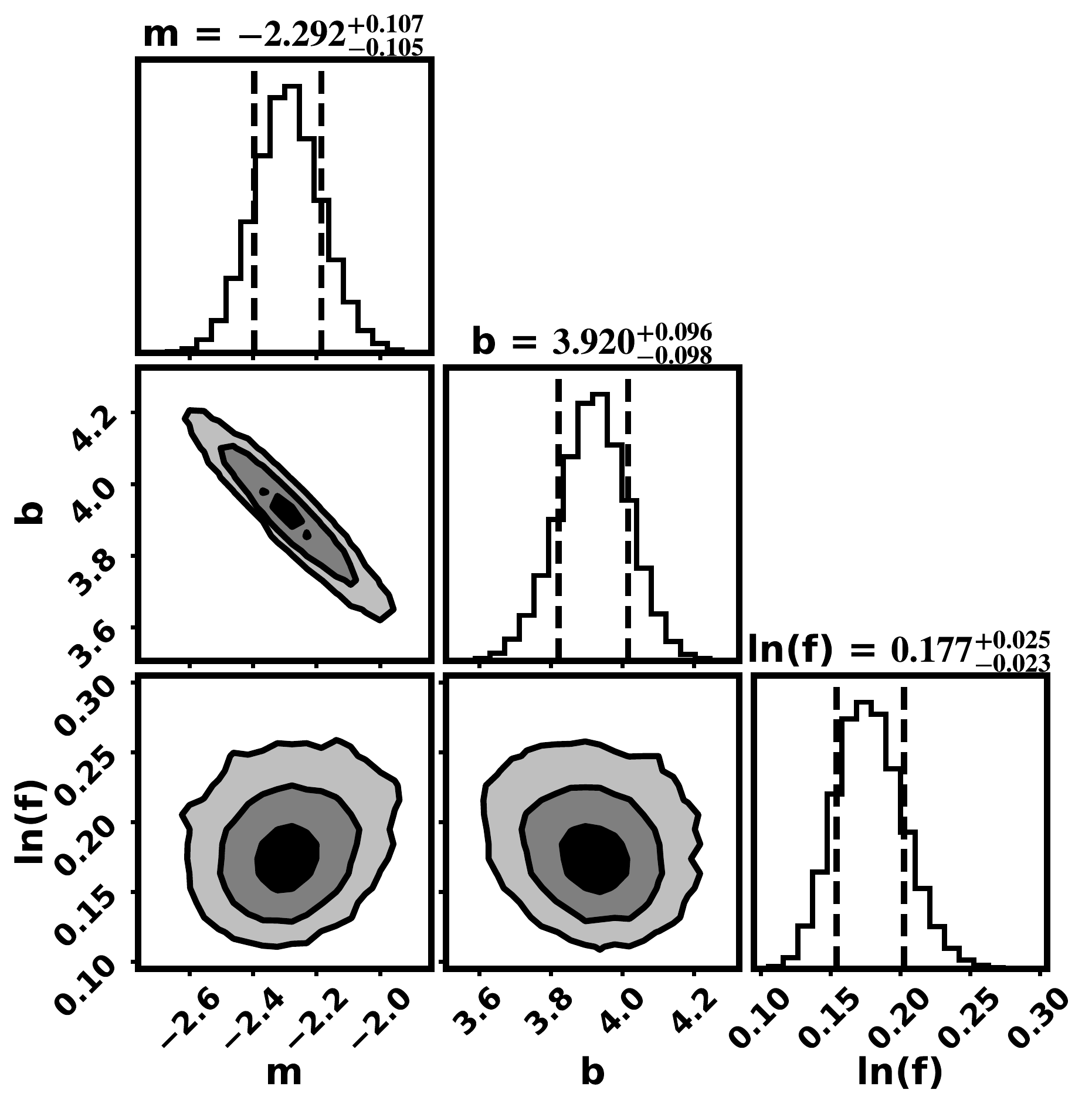}
    \includegraphics[width=0.32\textwidth]{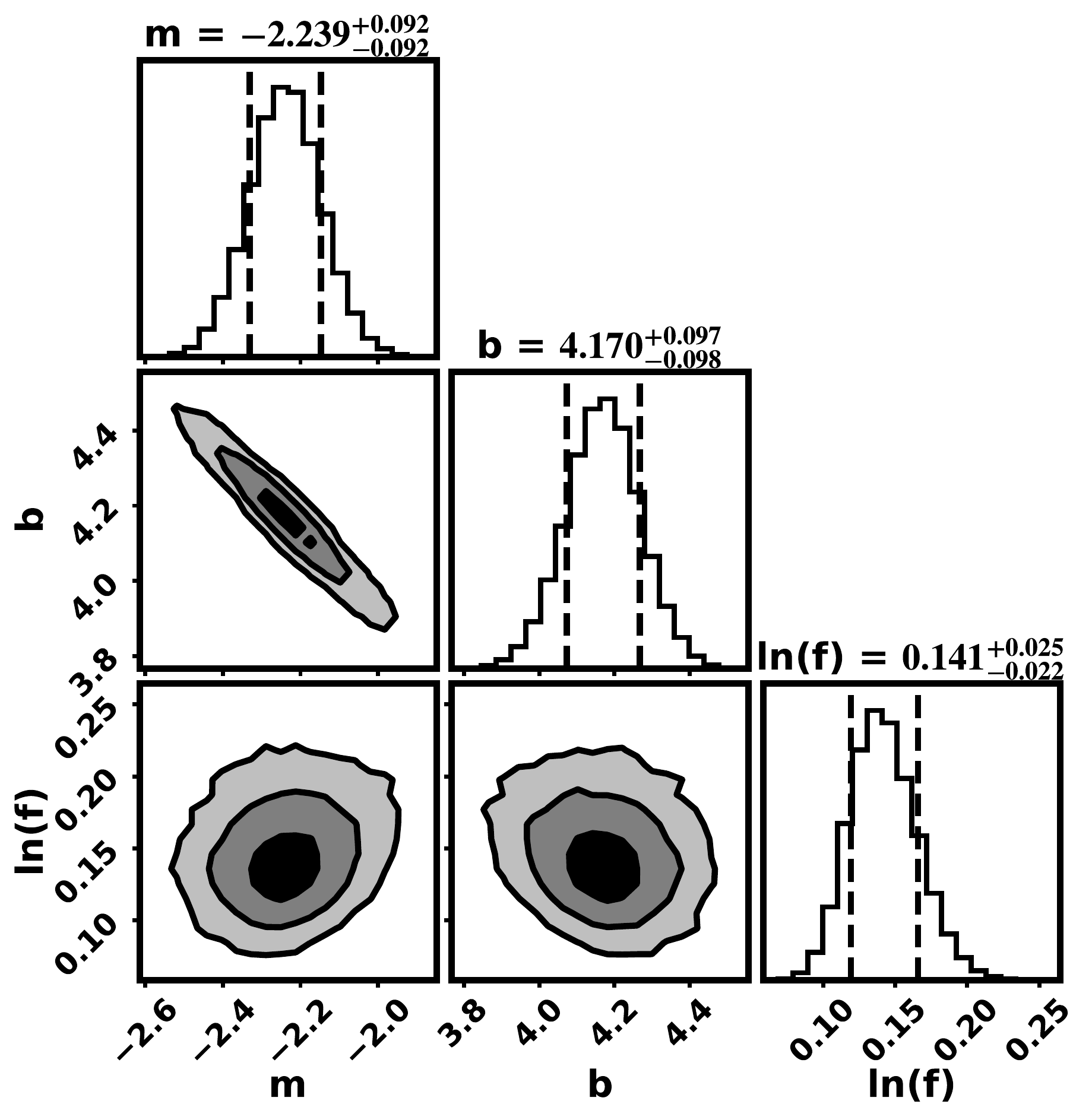}
    \caption{Corner plots of the parameters (slope, y-intercept, and missing uncertainty in the fit) from our MCMC model fits for \vgn\ (left), \vbn\ (center), and \vrn\ (right). The contour levels correspond to 1$\sigma$, 2$\sigma$, and 3$\sigma$ of the points (from darkest to lightest). The fit parameters are Gaussian distributed, with the expected covariance between the slope and y-intercept. Plot made using \texttt{corner} \citep{foreman2016corner}.}
    \label{fig:corner}
\end{figure*}

Figure \ref{fig:caliFits} shows the ages and \vn\ values for all three filters with the best-fit relation and random draws from the MCMC. All parameters were well constrained with Gaussian errors with the expected covariance between the slope and Y-intercept terms (Figure \ref{fig:corner}). The best-fit parameters and uncertainties for all filters are listed in Table \ref{tab:fitParams}.

\begin{deluxetable*}{lcccc}
\centering
\tabletypesize{\scriptsize}
\tablecaption{MCMC Fit Parameters \label{tab:fitParams}}
\tablehead{\colhead{Parameter} & \colhead{m} & \colhead{b} & \colhead{$\ln{f}$} & \colhead{$a$} }
\startdata
\vgn     & $-2.30\pm0.10$  & 3.928$^{+0.092}_{-0.093}$ & 0.178$^{+0.024}_{-0.022}$ & \nodata\\
\vbn     & $-2.29\pm 0.11$  & 3.920$^{+0.096}_{-0.098}$ & $0.177^{+0.025}_{-0.023}$ & \nodata\\
\vrn & $-2.239\pm0.092$ & $4.170^{+0.097}_{-0.098}$ & $0.141^{+0.025}_{-0.022}$ & \nodata \\
\vgn, $d$  & $-2.40 \pm 0.10$ & $4.22_{-0.14}^{+0.13}$ & $0.155_{-0.021}^{+0.025}$ & $-0.00112_{-0.00038}^{+0.00039}$\\
\vbn, $d$  &  $-2.461_{-0.098}^{+0.10}$ & $4.40_{-0.14}^{+0.13}$ & $0.129_{-0.022}^{+0.024}$ & $-0.00174 \pm 0.00037$ \\
\vrn, $d$ & $-2.376_{-0.082}^{+0.084}$ & $4.62 \pm 0.12$ & $0.089_{-0.021}^{+0.023}$ & $-0.00167_{-0.00032}^{+0.00031}$ \\
\enddata
\end{deluxetable*}

All three metrics followed a Skumanich-like decay ($\simeq t^n$) with age. Inverting $m$, we found $n$ varies from $-0.40$ to $-0.45$, consistent with the similar relation using full light curves \citep[$n=-0.37\pm0.16$;][]{Morris2020}. 

As can be seen in Figure~\ref{fig:caliFits}, the fit had a narrow range of solutions. The uncertainty in the output age from this relation was instead dominated by the $\ln{f}$ parameter. This implies a fundamental limit to the age precision of 14-18\% when using this technique. 

\subsection{Testing the relation}\label{sec:tests}
The significant $\ln{f}$ made clear that there are additional sources of variation in relation between \vn\ and age. The missing variation may be  related to the photometry (e.g., Poisson noise, \gaia's outlier rejection), assumptions about the input (e.g., inaccurate age uncertainties), and/or astrophysical effects (e.g., binarity and metallicity). Many of these cannot be studied in detail absent full light curves, but we explore some where we have the requisite data below. 

\textbf{Distance:} As seen in Figure \ref{fig:caliFits}, there is a tendency for more distant ($ \gtrsim 250$ pc) groups to sit below the fit and for the closest groups ($ \lesssim 125$ pc) to sit above the fit. The result is that more distant groups had an older variability-based age and closer groups a lower one. This may be due to the fact that more distant targets are (statistically) fainter, making it harder to detect the same level of variability in the presence of Poisson noise. 

We tested the effective distance ranges in all three filters. Removing the distant groups, $> 250$ pc, did not significantly change the calibration and all parameters agreed within the uncertainties. The decrease in $\ln{f}$ was insignificant. Similarly, removing the closest groups, $< 100$ pc, did not significantly effect the fit and all parameters agreed within uncertainties. 

We also explicitly fit a distance term of the form:
\begin{equation}
    \log_{10}({\rm{age}})\ \mathrm{[Myr]} = m\times Var_{90} + a\times d + b,
\end{equation}
where $d$ is the median distance (in parsecs) of the association members and $a$ is an additional fit parameter. The output parameters are included in Table \ref{tab:fitParams}. For the $G$-band, $a$ was consistent with 0 (2.9$\sigma$) but $a$ was significant the other two bands. The additional term suggests the inferred age shifts by about 0.1--0.2\% per pc in each filter. The correction thus becomes comparable to the intrinsic scatter in the relation for the most distant ($\gtrsim$ 300\,pc) or nearest ($\lesssim$100\,pc) groups. 

The fits accounting for distance had significantly lower $\ln{f}$ than those ignoring distance. For \vrn, the lower $\ln{f}$ suggested a limiting precision of 9\% (compared to 14\% when ignoring distance). For this reason, we suggest using the relations accounting for distance.

\textbf{Binaries:} High renormalised unit weight error \citep[RUWE;][]{GaiaDr2} values ($\gtrsim 1.2$) are often used to signify binary systems \citep{Pearce2019, Ziegler2020, Wood2022}. More restrictive RUWE cuts will not remove all binaries, but should remove enough of them to see if binaries have a significant impact on the result.

To test this, we added a RUWE cut of $< 1.3$ and an extreme cut of $< 1$. In both cases and for all filters, the $m$ and $b$ parameters agreed within $1 \sigma$. The $\ln{f}$ parameter for the $< 1.3$ cut agreed with our original fit, but increased by $>4\sigma$ for the $< 1$ cut. This may be because photometric variability can increase RUWE \citep{Belokurov2020}, as can the presence of a disk \citep{Fitton2022}. Thus, the tightest cut may be removing a subset of the most variable or youngest stars within a given population.

Individual \vgn\ values changed by $<1 \sigma$ after applying the RUWE $< 1.3$ cut for all groups except Taurus-Auriga, which varied by $3 \sigma$ (most likely due to a high fraction of members with disks). Additionally, $>$70\% of the \vgn\ values have smaller uncertainties before the RUWE $< 1.3$ cut was applied. We determine no RUWE cut is necessary, and applying one may negatively impact the resulting \vn\ value. 

\textbf{Field-star contamination:} There are often stars with motions and positions coincident with a group, particularly for the most diffuse populations. To explore this, we added stars from our field star sample (described in Section~\ref{sec:field}) to two groups and measured the effect on \vn. For this test, we used Lower Centaurus–Crux (17\,Myr) and Pleiades (112\,Myr). These were selected because together they span a range of ages and both groups have membership lists with low contamination rates. 

We added stars to each group from the field population randomly, only requiring that the added stars pass the same data quality and color cuts as the membership list. We then re-measured \vgn, as well as $Var_{G,50}$ and $Var_{G,75}$. 

The 90$^{th}$ percentile \varg\ value was least sensitive to interloper contamination (Figure~\ref{fig:contam}). Even at 30\% contamination level, field interlopers cause the median \varg\ to drop by about 20\%, while the 90$^{th}$ percentile value dropped by only 5\%. It took nearly a 75\% contamination level to drop \vgn\ by $\gtrsim$20\%. We conclude that field contamination had a weak effect on the result, which was a major motivation for selecting the 90$^{th}$ metric.

\textbf{Uncertainties from group size:} The limiting age precision from our relation is 9-16\% (when including distance) or 14-18\% (absent distance corrections). However, this ignores uncertainties in \vn\, which can be larger than the intrinsic uncertainty in the relation for low-mass groups.

To see how decreasing the sample size effects the final age uncertainty, we used the three largest calibration groups that span most of the age range: Lower Centaurus–Crux (17\,Myr), Psc-Eri (137\,Myr), and Praesepe (700\,Myr). We randomly removed stars from each group, recalculated the \vn\ (bootstrap) uncertainties, and propagated those to an uncertainty in age. We ignored uncertainties in the fit parameters and $\ln{f}$.

\begin{figure}[tb]
    \centering
    \includegraphics[width=0.47\textwidth]{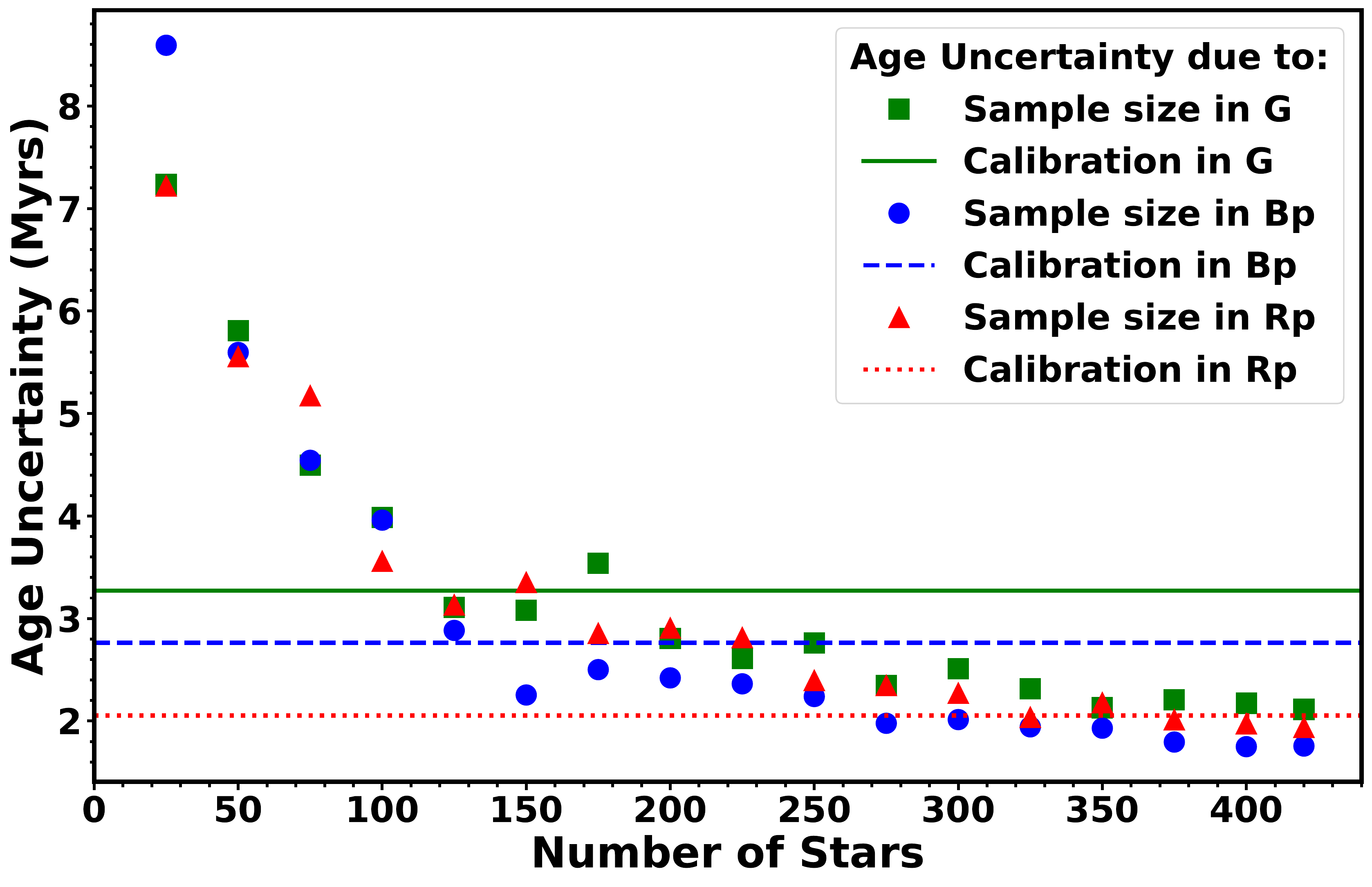}
    \includegraphics[width=0.47\textwidth]{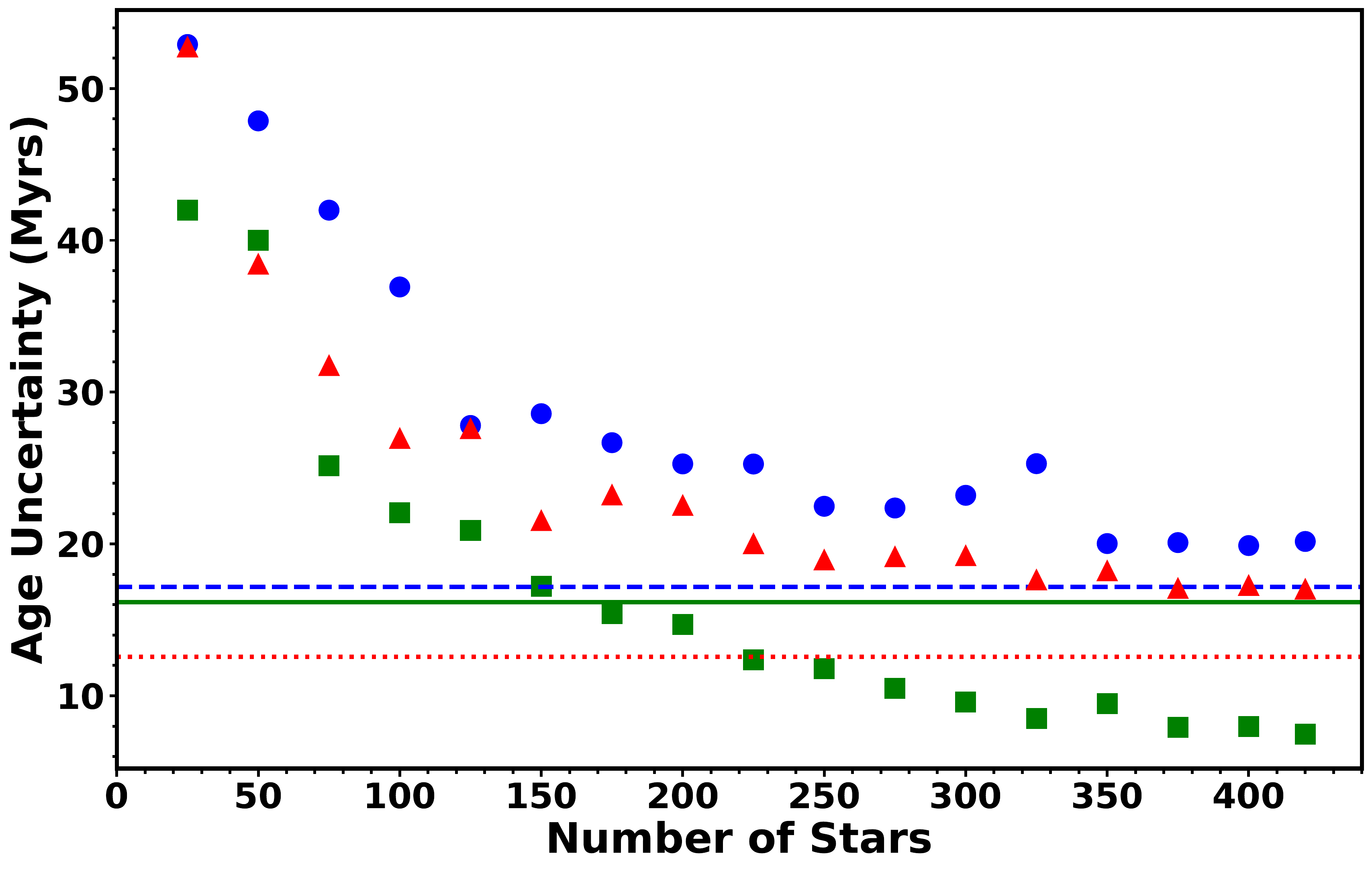}
    \includegraphics[width=0.47\textwidth]{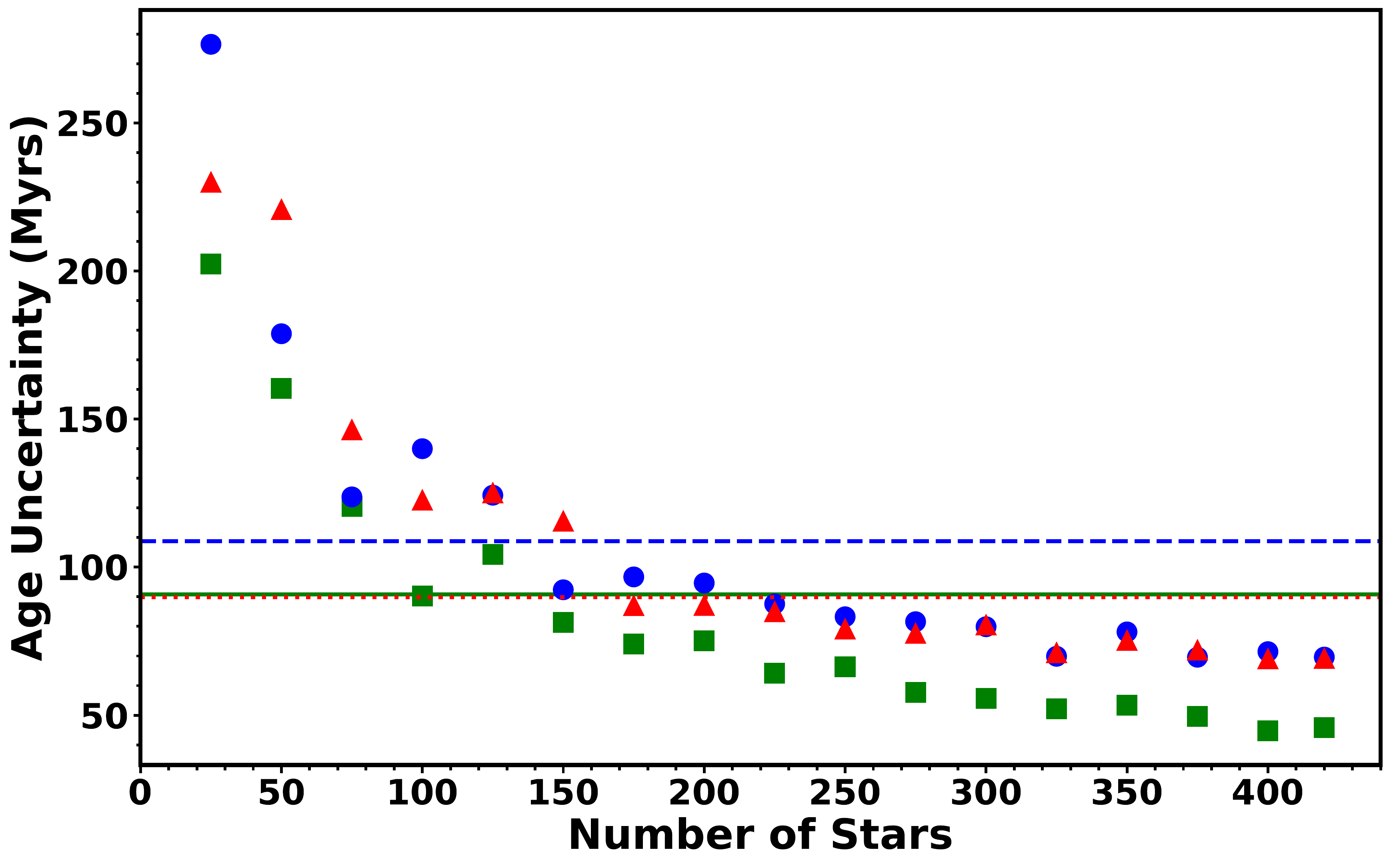}
    \caption{The effects of group size on the final age uncertainties. We used three groups of various ages, Lower Centaurus–Crux ($\sim$17 Myr, top), Psc-Eri ($\sim$130 Myr, middle), and Praesepe ($\sim$750 Myr, bottom). The individual points show the resulting age uncertainty arising from uncertainties in \vn, calculated by removing stars from these associations. The lines show the age uncertainty from the fit parameters' uncertainties (including $\ln{f}$). An optimal sample size would be where the uncertainty in \vn\ is below the fit uncertainty, which is dependent on the filter used but is typically $\sim$200-250 stars.}
    \label{fig:sizetest}
\end{figure}

As we show in Figure \ref{fig:sizetest}, uncertainties in \vn\ dominated the final age uncertainties for all bands and ages if the group has $\lesssim100$ stars (that pass all cuts). The effect was the strongest for Psc-Eri, where the age uncertainty from uncertainties in \vrn\ and \vbn\ do not drop below the calibration uncertainty even for samples $\gtrsim400$ stars. 

Figure~\ref{fig:sizetest} also makes clear that \vrn\ is not necessarily the best metric. While it has the smallest $\ln{f}$ value (Table~\ref{tab:fitParams}), the \vrn\ uncertainties are larger than those for \vgn, likely due to higher SNR in \gaia\ $G$ compared to $R_P$.

\textbf{Color cuts:} We included a color cut due to the metric becoming ineffective for mid-to-late M dwarfs (Figure~\ref{fig:colorseq}). To test the effect of this decision on the calibration, we reran the fit using stars with $B_P-R_P<3$ and again using stars with $B_P-R_P<2$. We found that the redder color cut had an insignificant effect on the fit parameters, but increased $\ln{f}$ by 2$\sigma$. As expected, the relation was diluted by the cooler M dwarfs where the metric is less effective. When using a bluer color cut, we found the fit parameters agreed with our original at 1$\sigma$, including $\ln{f}$. The main difference between the bluer cut and our original was that individual \vgn\ measurements had larger uncertainties due to the smaller sample of stars in each group.

\textbf{Impact of input age uncertainties:} Ages for the full sample were computed in an inhomogeneous way. This was unavoidable, as the methods used (and physics involved) to assign ages to older groups (e.g., main-sequence turn-off and asteroseismology of evolved stars) are subject to different systematics than methods that apply to younger stars (e.g., pre-main-sequence stars and lithium depletion boundary). Even in cases where the same method was used (e.g., CMD fitting), the choice of model and algorithm rarely matched between different analyses. Generally, ages for a given group agreed between source, but not necessarily the uncertainties. 

To explore the effect on the final relation, we reran the fit setting age uncertainties to zero. We found the fit parameters agree within 1$\sigma$; $\ln{f}$ increases marginally ($\simeq$1$\sigma$). If we instead assumed the calibration set age uncertainties are underestimated, $\ln{f}$ would be smaller. However, the change is insignificant; it dropped by only 1$\sigma$ when we doubled the input age uncertainties from those listed in Table~\ref{tab:sample}. To get a change of $\ge3\sigma$ in $\ln{f}$ required increasing input uncertainties by a factor of five. We conclude that our results are insensitive to our assumptions about the group age uncertainties.

There may be more complicated effects, such as systematic offsets in the ages based on the age or method. The complexity of these effects was too difficult to model in a robust way with the calibration sample here. However, we highlight that the output relation is only as good as the input ages. We also discuss how one could explore such effects in Section~\ref{sec:summary}.

\section{Application}\label{sec:application}
Here we highlight the utility of \vn\ and the age-\vn\ calibration by showing how they can be used to assess the assigned ages of newly identified groups, test if a young group is a real co-eval population instead of a collection of field stars with similar space velocities, and identify new young associations.

\subsection{Testing the ages of groups}\label{sec:testages}
We drew a collection of the Theia groups \citep{Kounkel2020} within 350pc that have at least 100 stars that pass the sample selection cuts (Section~\ref{sec:star}). In total, this included 59 groups with CMD-based ages from 16\,Myr to 2.6\,Gyr, comparable to our calibration sample. 

For each group, we calculated \vgn, \vbn, and \vrn, converted that to an age estimate in each filter, and took the weighted mean and uncertainty of the three ages. Combining the three age estimates may lead to underestimated uncertainties, as each fit was subject to some common systematics. However, the dominant uncertainty was due to scatter in \vn, and tests on the calibration sample suggested this simple combination was reasonable.

\begin{figure}[tbh]
    \centering
    \includegraphics[width=0.47\textwidth]{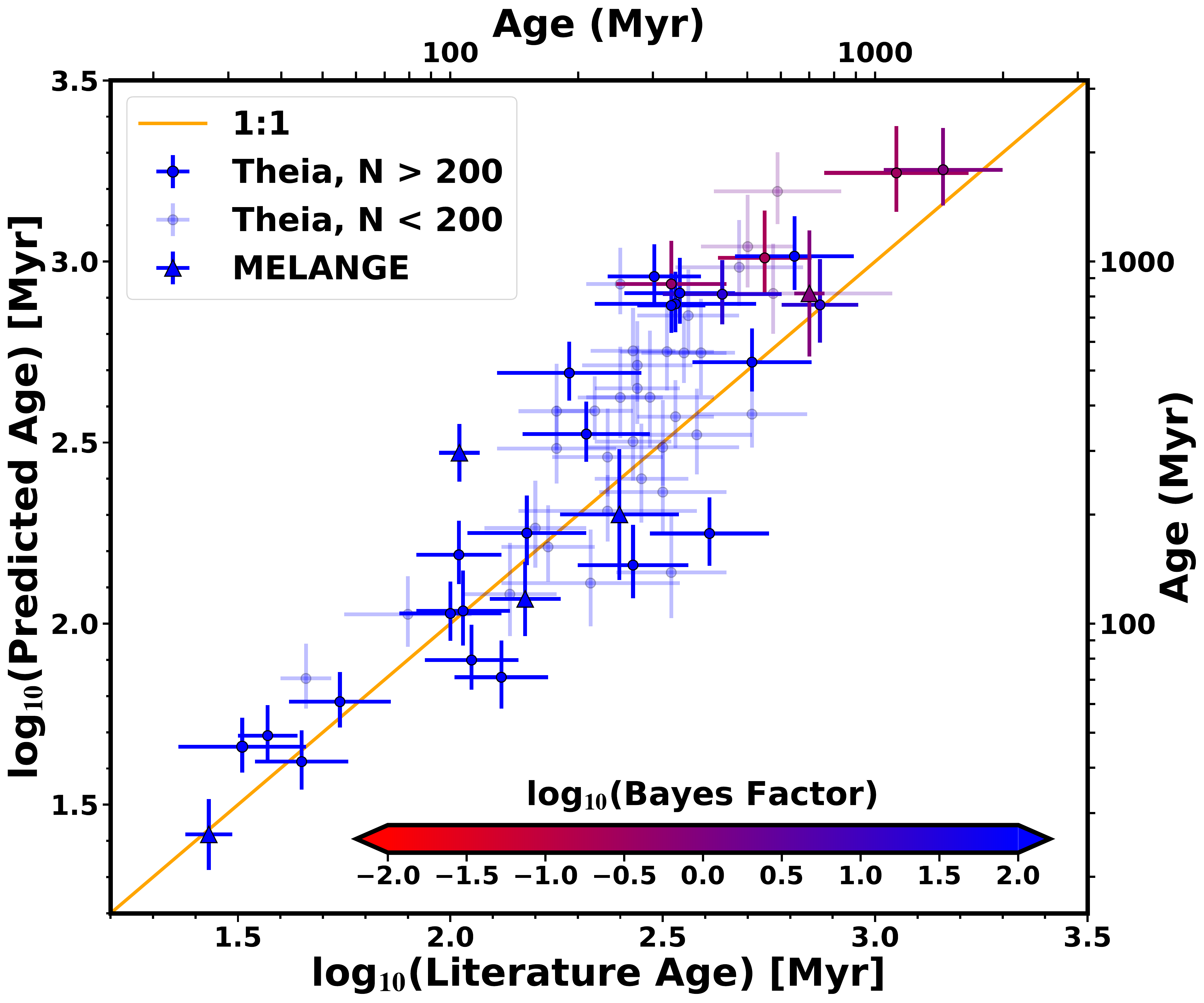}
    \caption{The predicted ages of the MELANGE groups (triangles) and Theia groups (circles) within 350pc and have $>100$ stars that pass sample cuts. The Theia groups with $> 100$ stars but $< 200$ stars are more transparent. A line showing the agreement is included for reference. Groups are colored by the Bayes factor comparing the probability of being a bona fide group or a collection of field stars (Equation~\ref{eqn:bayes}). A lower Bayes Factor (redder) indicates the association is more likely to be drawn from field stars. }
    \label{fig:theiaAges}
\end{figure}

Figure \ref{fig:theiaAges} compares our predicted ages to those from \citet{Kounkel2020}, determined using the neural network \texttt{Auriga}. \texttt{Auriga} uses quantities derived from the photometry and parallaxes (the CMD), such as the ratio of high and low mass stars and the ratio of post-, main-, and pre-sequence stars. 

Of the 59 groups, 48 (80\%) have variability ages 3$\sigma$ consistent with those from \cite{Kounkel2020}. Of the 11 discrepant groups, 8 are $\gtrsim$300\,Myr with variability ages significantly higher than the Auriga-determined age. This can be seen in Figure~\ref{fig:theiaAges} as an overdensity of points in the top half of the age distribution sitting above the 1:1 line. Below $\gtrsim$300\,Myr, there are a similar number of points on either side of the 1:1 line. 

The systematic offset at older ages is, in part, because \vn\ is more effective at younger ages. Another factor is likely the \texttt{Auriga} ages. \citet{Kounkel2020}, comparing \texttt{Auriga} ages to those from the literature, found that \texttt{Auriga} tends to overestimate ages for groups $>300$\,Myr and underestimate ages for younger ones. This roughly matches our own comparison. It is also possible that some older Theia groups are field stars with coincident space motions, which we discuss in the next section.

Most of the variability-based ages are {\it more precise} than the isochronal ages, particularly at young ages. Of the 48 groups where the two ages agree, 26 (55\%) have variability-based age uncertainties below the Auriga-based age uncertainties. For groups $<100$\,Myr, where \vn\ works best, five of seven (70\%) have smaller age uncertainties when using variability ages compared to the \texttt{Auriga} ages. 

We performed a similar test on the five published MELANGE groups. All but one predicted ages agreed within 1$\sigma$ to their reported values. The exception, MELANGE-3 had a 3.5$\sigma$ older variability age ($\simeq$300\,Myr) compared to the age derived from lithium and rotation \citep[105\,Myr; ][]{Barber2022}. This may have been because the group lands at the distance limit of our calibration sample (326\,pc) and has a high field contamination rate \citep[$\simeq$50\%;][]{Barber2022}. 

All associations we tested are listed in Table~\ref{tab:testsample}, including the literature age and variability-based age.

\subsection{Testing the validity of a group}\label{sec:validity}

Automated machine-learning tools designed to find overdensities of stars \citep[e.g., HDBSCAN; ][]{McInnes2017} run the risk of identifying collections of stars with similar velocities that are neither bound nor co-eval. Our results in Section~\ref{sec:testages} hint at this problem; there are Theia groups with variability ages higher than the CMD-based age, and many of these groups have variability ages similar to what we expect when drawing random field stars ($\sim$1\,Gyr since we are using the 90th percentile of \var). 

Groups with variability levels closer to the local field stars than the values predicted by their age are unlikely to be real co-eval populations. We quantified this using a Bayes factor:
\begin{equation}\label{eqn:bayes}
    K = \frac{P(Var_{90}|G)}{P(Var_{90}|F)},
\end{equation}
where $P(Var_{90}|G)$ is the probability of measuring the \vn\ value given that the stars are drawn from a real population (with an assumed age), and $P(Var_{90}|F)$ is the probability assuming stars are drawn from the field. 

We computed both terms assuming Gaussian distributions. We restricted our analysis to \vgn, although the other bands gave similar results. The numerator term we calculated by propagating the assigned age into a predicted \vgn\ and uncertainty (accounting for age and fit uncertainties). For the denominator, we drew a random sample of stars, matching the group size, with distances within 0.1 mas of the group distance and satisfying all cuts from Section~\ref{sec:star}. We list the resulting $K$ values for each group in Table~\ref{tab:testsample}.

Following the Jeffreys' scale \citep{jeffreys1961theory, kass1995bayes}, we adopt a threshold of $|\log_{10}(K)|<0.5$ (approximately 3-to-1 odds) as the threshold for substantial evidence. Four (of five) MELANGE groups and 53 (of 59) Theia groups we tested had substantial evidence of being a real association ($\log_{10}(K) > 0.5$). Four of the remaining Theia groups (Theia 514, 793, 1098, and 1532) and the one MELANGE group (MELANGE-2) were ambiguous ($-0.5 < \log_{10}(K) < 0.5$). These were cases where the variability was consistent with a field population, but the CMD age was also relatively old. MELANGE-2 is also the smallest group (32 stars), making this test challenging. The remaining two Theia groups (Theia 810 and 1358) have substantial evidence for not being a real association ($\log_{10}(K) < -0.5$). Using a more definitive cut of $|\log_{10}(K)|>2.2$ moves 9 groups, including Theia 810 and 1358, into the ambiguous category.

Consistent with our findings in the previous section, all seven of ambiguous and unlikely groups are $>300$\,Myr and have variability ages above their CMD-based age, helping to explain the excess of points above the 1:1 line in Figure~\ref{fig:theiaAges}.

\subsection{Finding new associations}

\begin{figure*}[tbh]
    \centering
    \includegraphics[width=0.97\textwidth]{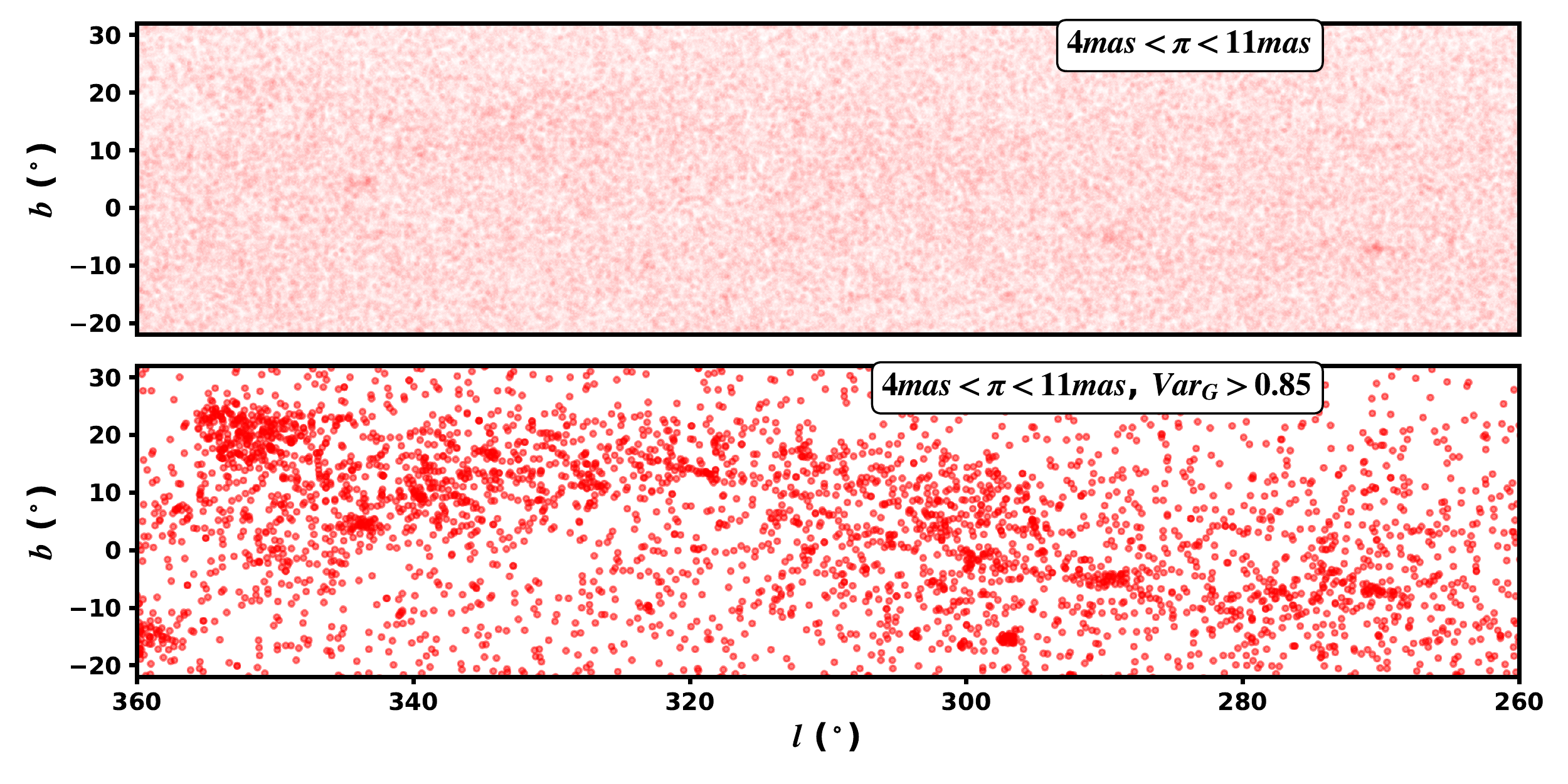}
    \caption{Stars in the region around the Sco-Cen OB association. The top panel shows all the stars with $4<\pi<11$, an extremely generous cut that includes nearly all of Sco-Cen. The bottom shows the same parallax cut, but adding a requirement that the star is in the top 2\% of \varg. Some of the youngest regions (e.g., Upper Scorpius in the top-left and Corona Australis in the bottom left) show quite clearly after a simple variability cut, as well as dense sub-groups like Lupus. }
    \label{fig:scocen}
\end{figure*}

In Figure \ref{fig:scocen}, we can see potential of \var\ for searching for new associations. We first show all stars within the general area of Scorpius-Centarus ($4<\pi<11$) and satisfying the cuts from Section~\ref{sec:star}. A few of the denser regions show up, but not the overall structure. However, when we only include stars in the top 2\% of \varg, the Sco-Cen population is clear. Further, many of the youngest groups (e.g., Corona Australis and Upper Scorpius) are the most prominent after applying the \varg\ cut. 

One could have made a similar or better Sco-Cen member selection using \gaia\ astrometry or CMD position. However, the benefit was that we were able to identify Sco-Cen and numerous sub-populations from excess noise in the \gaia\ photometry {\bf alone}. This would have worked even without a parallax cut; we only applied that to keep the sample size reasonable. One could therefore combine \var\ with positional, kinematic, and other age information to identify groups that are far more diffuse or otherwise challenging to identify and confirm purely from the traditional positional and kinematic information.

\section{Summary and Conclusions}\label{sec:summary}

\subsection{Summary of findings}
Earlier work from \citet{Guidry2021} and \citet{Barlow2022} showed that one can use the excess flux uncertainty from \gaia\ to identify variable white dwarfs and hot subdwarfs, respectively. Here we have extended this work to young stars. Specifically, we 1) modified the excess uncertainty metric using the median flux uncertainties provided by \gaia\ \citep{Riello2021_phot}, 2) showed that our new metric (\var) scales with age for FGK and early M dwarfs, 3) calibrated the relation between the 90th-percentile of \var\ (\vn) and age, 4) demonstrated how the metric can be used to estimate the ages of young populations, confirm which young populations are real, and search for new young groups. 

Our results confirmed a correlation between stellar variability and age. Our calibrations, in all bands, whether or not we included distance corrections, yield a Skumanich decay with age consistent with similar relations using full light curves \citep{Morris2020}. 

We found a narrow range of solutions in our calibration, and the uncertainty of the output is dominated by $\ln{f}$. This suggests the scatter in the relation was astrophysical and the fundamental age precision limit using the variability-age relations is $\ge$9\%. 

The methods described here work best on populations below $<500$\,Myr and those with $>$100 stars. Testing if a group is real or looking for new groups both work better at the youngest ages. For the former, the probability of drawing a population of highly variable stars by chance is negligibly low. For the latter, at $\simeq100$\,Myr late-type stars show \var\ levels well above the field population.

We provide a copy of our software that can be used to compute \var\ and \vn\footnote{\url{https://github.com/madysonb/eva}}.

\subsection{Are the Theia Strings real structures?}

\citet{Kounkel2019} constructed Theia strings by manually combining sets of groups (originally identified by HDBSCAN) with similar ages and coherent spatial and kinematic structure. \citet{Zucker2022b} argued that the individual groups that make the strings may be real populations but were unlikely to be part of a single bona fide structure. They primarily pointed that each string has a high velocity dispersion, yielding a high virial mass and breakup timescales much shorter than the group ages. \cite{Manea2022} found a majority of Theia structures contain abundances more homogeneous than their local fields, noting that of the 10 strings and 8 compact groups tested, Theia 1415 was the only string (and group) they found to have a high abundance dispersion more closely matching local background stars. However, \citet{Zucker2022b} argued this could happen even by chance if many of the sub-components of the string are young populations and does not require them to be part of a larger structure. 

Our results contrast with \citet{Zucker2022b}. We found the majority of Theia groups contain variability measurements consistent with their reported isochronal ages. Just considering the strings, the two age estimates matched in 36 of 45 cases (80\%). It is unlikely these numbers would match so often if each string were comprised of many groups with varying ages. Exactly how unlikely depends on the age and age spread between subgroups, but if we assume purely random draws from the Theia group ages, then we would expect no matches over the 45 strings by chance alone. Further, most Theia strings passed our validity test. Only one of the strings (Theia 830) has substantial evidence for not being a real association. Even assuming cases with ambiguous results (similar probability of being pulled from the field versus a real group) were not real, 39 of 43 (90\%) of the Theia strings had substantial evidence of being real. 

These results could be reconciled if some of the sub-groups are associated and some are not and/or the strings contain some field contamination. Because \vn\ is weakly impacted by contamination (Figure~\ref{fig:contam}), most of the sub-groups within a string could be un-associated and we would still get an age consistent with the CMD-based age. However, \citet{Zucker2022b} would find a high velocity dispersion even if just a few of the sub-groups were disconnected. If the unassociated groups were preferentially not real (field stars) or older than the main group, this may also help explain why, among the $>300$\,Myr groups, the variability ages were preferentially higher than the isochronal ages (Figure~\ref{fig:theiaAges}). 

A similar explanation is that each string is composed of multiple populations with similar but not identical ages and kinematics. An example is the Sco-Cen OB association, which is comprised of at least three, but probably many more populations \citep[e.g.][]{Kerr2021, Luhman2022}. These sub-groups are unbound and have slightly differing kinematics and ages \citep{Wright2018}. The velocity difference between parts of Sco-Cen can exceed 10\,\kms\ \citep{Zerjal2023}, similar to many of the Theia strings, and this dispersion would only grow with time. Sco-Cen would have broken apart by the age of the oldest Theia strings, but many strings are $<50$\,Myr and a denser equivalent of Sco-Cen may still show up for hundreds of millions of years. 

Our results are closer in line with that of \citet{Hunt2023}. In their all-sky search, they recovered most of the Theia groups (including the strings) out to ages of $\simeq100$\,Myr, but almost none above 1\,Gyr. Similarly, all the $<100$\,Myr groups passed our viability test and have Auriga ages consistent with our own. Rejected/ambiguous cases tended to be $>1$\,Gyr. This is, in part, because our metric works best on young groups and groups are easiest to distinguish from (mostly old) field stars when young. However, this fits with a scenario where the youngest groups are robust while the older ones contain a mix of real populations and random stellar overdensities.

\subsection{Benefits of Var}
Our age-\vn\ calibration can yield ages with $\simeq$10\% precision, provided the population has a sufficient number of FGK and early M star members ($\gg$100). This is competitive with other methods, like isochrone fitting. We can see this in the comparison of our variability ages to the CMD-based ages from \cite{Kounkel2020}; variability-based ages were often more precise, particularly below $<$200\,Myr. 

A major benefit of this method is the limited information needed. We are able to get quick age estimates using available \gaia\ DR3 data and without the need for collecting additional rotation period and lithium measurements. For example, the age for MELANGE-4 is based on Lithium absorption, which requires multiple nights of observations \citep{Wood2022}. While not as precise, we calculated a similar age from just \gaia\ data alone ($26^{+8}_{-5}$\,Myr compared to $27\pm3$\,Myr). The \vn\ calibration is also independent of CMD- or abundance-based methods, meaning it can be combined to improve precision.

Another example is MELANGE-1, which was identified using \texttt{FriendFinder} \citep{THYMEV}, by selecting stars with similar positions and motions to a given target. The population showed weak evidence of spatial or kinematic over densities, and required additional radial velocity, rotation periods, and Lithium measurements to confirm the group is real and measure its age. As we showed in Section~\ref{sec:application}, we obtained a consistent (but less precise) age and confirmed its a real population from \gaia\ data alone. 

While other fitting methods, such as isochrone fittings, rely on the distribution of pre-, post-, and main-sequence stars, \vn\ works in age ranges where there are few or no pre-main sequence or evolved stars (approximately 200-500\,Myr). It is also independent of extinction (provided the stars are sufficiently bright). Lastly, the method will grow in effectiveness as \gaia\ collects additional data and we can calibrate past 350\,pc. 

While individual stars cannot be aged using this method (see Figure~\ref{fig:groupsHistograms}), \var\ can be used as another metric for identifying high probability group members. This is especially useful for diffuse groups with are few pre-main sequence stars (e.g., AB Dor). \var\ can be used to identify candidate young stars in the field, and other methods can be used to confirm membership.

\subsection{Limitations}
The most obvious place where \vn\ failed was MELANGE-3; \vn\ suggested an age of 300$\pm$60\,Myr but the rotation, lithium levels, and CMD all indicate an age of 105$\pm10$\,Myr \citep{Barber2022}. This discrepancy also stands out because the other disagreements for MELANGE and Theia ages were for $>500$\,Myr, where \vn\ is less effective and groups are harder to distinguish from the field (less likely to be real structures). MELANGE-3 passed our validity test. 

For the best age estimates, the metric requires a sample sizes of at least 100, while ages can be derived from a CMD with a handful of turn-off or pre-main-sequence stars. Turn-off stars are also available at far greater distances than 350\,pc. The size limitation is also a problem for low-mass nearby groups like those from \citet{Moranta2022}, the majority of which have fewer than 100 members. 

 \var\ is ineffective for stars cooler than $\simeq$M3V (Figure~\ref{fig:colorseq}). We suspect this is a mix of a few effects; 1) the \gaia\ fitted uncertainties are calibrated mostly on FGK stars and do not include a color term, 2) M dwarfs are intrinsically fainter than FGK dwarfs so the distance effect (brightness) discussed in Section~\ref{sec:tests} is stronger, 3) M dwarfs are variable for longer and their variation may saturate below 100-500\,Myr \citep{Jackson2012, Kiman2021}.

\subsection{Future work}
The methods described here could be used to test which sub-groups of a  given Theia string are co-eval. For example, we could see if the \var\ distributions for each sub-group are consistent with being drawn from the same parent population (one single-aged group). A more complex mixture model would also let us test if the strings are consistent with a mix of a young population and field contaminants or multiple young populations. This could be done in conjunction with analysis of the kinematics and position (e.g., which groups combine together to yield a low velocity dispersion while maintaining a consistent \var\ distribution). 

In Section~\ref{sec:tests}, we explored the effects of over- or under-estimated age uncertainties. A more complicated concern is systematics in the ages (or uncertainties) based on the age, method, or model used. One way to test our sensitivity to this would be to split the calibration sample up by method, drop out groups using a common method, and redo the calibration. The problem is that almost all $>500$\,Myr groups have their ages from isochrones, but with significant variations in the algorithm, model, or stars. Thus, proper treatment would involve re-doing some of the original age estimates and careful decision-making about what counts as a common method.

It may be possible to recover \var\ as a useful metric for mid-to-late M dwarfs. One option is to re-calibrate \gaia\ photometric uncertainty estimates including color as a parameter. Similarly, one could compare \var\ to the expected uncertainty for a set of stars of similar distance, brightness, and $B_P-R_P$ color. This would effectively change \var\ to a metric that compares the photometric uncertainties to that of the median star of similar spectral type and apparent brightness. 

The number of associations with high-quality membership lists and well-determined ages decreases significantly past 350\,pc. This made it challenging to calibrate the relation further. The reach of \gaia\ data and new search tools are expanding the list of groups \citep[e.g.][]{Qin2022,He2022}. More complete membership lists and more detailed age estimates for these groups would be invaluable to calibrate Equation~\ref{eqn:fit} to 500\,pc or beyond. 

Another route for improvement would be to use the $BPRP$ spectra from \gaia. Stellar variability is stronger in some parts of the spectrum than others (e.g., around H$\alpha$). One could create synthetic photometry from the spectra \citep{GaiaCollaboration2022} tuned to these wavelength regions, which should be more effective than broadband photometry alone.

% \acknowledgements
\begin{acknowledgements}
The authors thank the anonymous referee for their thoughtful and detailed comments. We thank Halee and Bandit for their tireless efforts to interrupt Zoom meetings between the two authors. We also thank Marina Kounkel and Pa Chia Thao for their comments on the manuscript, and the UNC Journal Club for discussing the \citet{Guidry2021} paper, which spawned the idea for this work. 

MGB and AWM were both supported by a grant from the NSF CAREER program (AST-2143763) and a grant from NASA's exoplanet research program (XRP 80NSSC21K0393). 

This research has made use of the tool provided by Gaia DPAC to reproduce the Gaia (E)DR3 photometric uncertainties described in the GAIA-C5-TN-UB-JMC-031 technical note using data in Riello et al. (2021).

This work presents results from the European Space Agency (ESA) space mission Gaia. Gaia data are being processed by the Gaia Data Processing and Analysis Consortium (DPAC). Funding for the DPAC is provided by national institutions, in particular the institutions participating in the Gaia MultiLateral Agreement (MLA). The Gaia mission website is https://www.cosmos.esa.int/gaia. The Gaia archive website is https://archives.esac.esa.int/gaia.

\end{acknowledgements}

\facilities{\gaia}

\software{\texttt{emcee}, \texttt{corner.py}, matplotlib \citep{Hunter2007}, \texttt{Astropy} \citep{Astropy2013, AstropyCollaboration2018}, \texttt{numpy} \citep{harris2020}, \texttt{scipy} \citep{Virtanen2020}.}

\startlongtable
\begin{deluxetable*}{lclccccc}%% 
\centering
\tabletypesize{\scriptsize}
\tablecaption{Test Group Results  \label{tab:testsample}}
\tablehead{\colhead{Name} & \colhead{Literature Age\tablenotemark{a}} & \colhead{Age} & \colhead{Variability Age} & \colhead{Distance\tablenotemark{b}} & \colhead{N$_{\rm{stars}}$\tablenotemark{c}} & \colhead{Bayes Factor} & \colhead{String?}
\\
\colhead{} & \colhead{(Myr)} & \colhead{Reference} & \colhead{(Myr)} & \colhead{pc}  & \colhead{} & \colhead{$\log_{10}$(K)} & \colhead{Y/N} }
\startdata
Theia 44 & $32_{-9}^{+13}$ & \citet{Kounkel2020} & $46_{-7}^{+9}$ & 127 & 109 & 27.4 & Y \\
Theia 115 & $45_{-10}^{+13}$ & \citet{Kounkel2020} & $41_{-7}^{+9}$ & 178 & 202 & 62.0& Y \\
Theia 116 & $55_{-13}^{+17}$ & \citet{Kounkel2020} & $61_{-9}^{+13}$ & 226 & 599 & 221.6& Y \\
Theia 120 & $37_{-6}^{+6}$ & \citet{Kounkel2020} & $49_{-8}^{+11}$ & 327 & 427 & 177.8 & Y \\
Theia 138 & $46_{-6}^{+7}$ & \citet{Kounkel2020} & $70_{-12}^{+17}$ & 359 & 189 & 67.5 & Y \\
Theia 160 & $79_{-23}^{+33}$ & \citet{Kounkel2020} & $106_{-20}^{+29}$ & 175 & 139 & 24.9 & Y \\
Theia 163 & $100_{-24}^{+32}$ & \citet{Kounkel2020} & $106_{-17}^{+24}$ & 318 & 474 & 136.5 & Y \\
Theia 164 & $112_{-25}^{+32}$ & \citet{Kounkel2020} & $80_{-14}^{+20}$ & 314 & 203 & 71.8 & Y \\
Theia 211 & $275_{-71}^{+96}$ & \citet{Kounkel2020} & $515_{-107}^{+164}$ & 215 & 141 & 6.4 & N \\
Theia 214 & $158_{-38}^{+50}$ & \citet{Kounkel2020} & $184_{-40}^{+65}$ & 218 & 144 & 28.3 & Y \\
Theia 215 & $132_{-29}^{+38}$ & \citet{Kounkel2020} & $71_{-13}^{+19}$ & 231 & 268 & 79.4 & Y \\
Theia 216 & $107_{-24}^{+31}$ & \citet{Kounkel2020} & $108_{-21}^{+32}$ & 230 & 220 & 47.3 & Y \\
Theia 219 & $219_{-41}^{+50}$ & \citet{Kounkel2020} & $384_{-65}^{+93}$ & 262 & 182 & 14.8 & Y \\
Theia 227 & $191_{-62}^{+91}$ & \citet{Kounkel2020} & $493_{-79}^{+109}$ & 329 & 764 & 37.6 & N \\
Theia 228 & $138_{-31}^{+40}$ & \citet{Kounkel2020} & $119_{-28}^{+48}$ & 329 & 102 & 22.9 & Y \\
Theia 303 & $151_{-42}^{+58}$ & \citet{Kounkel2020} & $177_{-32}^{+48}$ & 224 & 226 & 36.1 & Y \\
Theia 311 & $209_{-61}^{+86}$ & \citet{Kounkel2020} & $337_{-56}^{+78}$ & 288 & 264 & 23.9 & Y \\
Theia 370 & $214_{-82}^{+133}$ & \citet{Kounkel2020} & $129_{-31}^{+53}$ & 146 & 132 & 12.8 & N \\
Theia 430 & $178_{-49}^{+68}$ & \citet{Kounkel2020} & $305_{-61}^{+95}$ & 161 & 117 & 6.7 & Y \\
Theia 431 & $234_{-90}^{+146}$ & \citet{Kounkel2020} & $205_{-36}^{+53}$ & 167 & 176 & 20.0 & Y \\
Theia 433 & $347_{-90}^{+121}$ & \citet{Kounkel2020} & $820_{-141}^{+201}$ & 235 & 255 & 5.4 & Y \\
Theia 438 & $316_{-92}^{+130}$ & \citet{Kounkel2020} & $230_{-51}^{+84}$ & 263 & 106 & 12.6 & Y \\
Theia 506 & $407_{-112}^{+155}$ & \citet{Kounkel2020} & $177_{-32}^{+47}$ & 93 & 202 & 18.3 & Y \\
Theia 509 & $269_{-70}^{+94}$ & \citet{Kounkel2020} & $146_{-29}^{+42}$ & 145 & 240 & 21.3 & Y \\
Theia 514 & $331_{-86}^{+116}$ & \citet{Kounkel2020} & $863_{-177}^{+267}$ & 282 & 223 & -0.3 & N \\
Theia 515 & $275_{-57}^{+71}$ & \citet{Kounkel2020} & $450_{-92}^{+139}$ & 297 & 125 & 6.0 & N \\
Theia 516 & $269_{-50}^{+62}$ & \citet{Kounkel2020} & $318_{-70}^{+111}$ & 300 & 133 & 8.4 & Y \\
Theia 519 & $389_{-65}^{+79}$ & \citet{Kounkel2020} & $564_{-132}^{+223}$ & 343 & 120 & 2.7 & N \\
Theia 595 & $513_{-141}^{+195}$ & \citet{Kounkel2020} & $527_{-90}^{+125}$ & 113 & 322 & 3.4 & Y \\
Theia 599 & $302_{-68}^{+87}$ & \citet{Kounkel2020} & $906_{-150}^{+207}$ & 238 & 372 & 2.8 & Y \\
Theia 600 & $269_{-55}^{+70}$ & \citet{Kounkel2020} & $569_{-113}^{+177}$ & 260 & 143 & 6.8 & N \\
Theia 603 & $234_{-61}^{+82}$ & \citet{Kounkel2020} & $288_{-64}^{+105}$ & 287 & 110 & 7.3 & Y \\
Theia 605 & $105_{-22}^{+27}$ & \citet{Kounkel2020} & $154_{-25}^{+38}$ & 319 & 278 & 56.0 & Y \\
Theia 678 & $339_{-120}^{+186}$ & \citet{Kounkel2020} & $767_{-125}^{+174}$ & 144 & 427 & 2.9 & Y \\
Theia 683 & $295_{-86}^{+122}$ & \citet{Kounkel2020} & $421_{-114}^{+214}$ & 216 & 132 & 2.6 & Y \\
Theia 684 & $355_{-73}^{+92}$ & \citet{Kounkel2020} & $560_{-97}^{+143}$ & 216 & 141 & 5.3 & Y \\
Theia 685 & $331_{-56}^{+67}$ & \citet{Kounkel2020} & $758_{-122}^{+166}$ & 229 & 369 & 6.9 & Y \\
Theia 695 & $178_{-33}^{+41}$ & \citet{Kounkel2020} & $385_{-84}^{+136}$ & 304 & 104 & 9.4 & Y \\
Theia 786 & $282_{-63}^{+81}$ & \citet{Kounkel2020} & $253_{-61}^{+104}$ & 152 & 118 & 8.0 & Y \\
Theia 790 & $324_{-72}^{+93}$ & \citet{Kounkel2020} & $563_{-124}^{+203}$ & 200 & 107 & 3.0 & N \\
Theia 792 & $339_{-63}^{+78}$ & \citet{Kounkel2020} & $372_{-68}^{+100}$ & 211 & 142 & 8.2 & Y \\
Theia 793 & $501_{-112}^{+144}$ & \citet{Kounkel2020} & $1110_{-257}^{+420}$ & 241 & 109 & 0.4 & Y \\
Theia 796 & $251_{-52}^{+65}$ & \citet{Kounkel2020} & $425_{-98}^{+158}$ & 229 & 109 & 4.3 & Y \\
Theia 801 & $316_{-107}^{+162}$ & \citet{Kounkel2020} & $307_{-67}^{+107}$ & 260 & 114 & 12.4 & N \\
Theia 807 & $170_{-38}^{+49}$ & \citet{Kounkel2020} & $164_{-33}^{+49}$ & 311 & 119 & 24.7 & N \\
Theia 809 & $251_{-42}^{+51}$ & \citet{Kounkel2020} & $868_{-153}^{+227}$ & 311 & 140 & 2.0 & Y \\
Theia 810 & $550_{-123}^{+158}$ & \citet{Kounkel2020} & $1030_{-222}^{+354}$ & 345 & 227 & -0.6 & Y \\
Theia 906 & $437_{-120}^{+166}$ & \citet{Kounkel2020} & $810_{-141}^{+201}$ & 128 & 266 & 1.4 & Y \\
Theia 907 & $575_{-273}^{+521}$ & \citet{Kounkel2020} & $819_{-185}^{+297}$ & 148 & 125 & 0.6 & N \\
Theia 908 & $646_{-178}^{+246}$ & \citet{Kounkel2020} & $1035_{-196}^{+294}$ & 214 & 298 & 2.0  & Y \\
Theia 912 & $479_{-140}^{+197}$ & \citet{Kounkel2020} & $959_{-209}^{+339}$ & 262 & 120 & 1.2 & Y \\
Theia 1007 & $513_{-133}^{+179}$ & \citet{Kounkel2020} & $380_{-73}^{+109}$ & 169 & 182 & 7.1 & Y \\
Theia 1008 & $363_{-88}^{+116}$ & \citet{Kounkel2020} & $714_{-153}^{+243}$ & 196 & 181 & 4.0 & Y \\
Theia 1010 & $380_{-98}^{+133}$ & \citet{Kounkel2020} & $333_{-73}^{+114}$ & 241 & 135 & 13.8 & N \\
Theia 1012 & $331_{-86}^{+116}$ & \citet{Kounkel2020} & $140_{-35}^{+61}$ & 298 & 127 & 26.7 & Y \\
Theia 1094 & $741_{-139}^{+171}$ & \citet{Kounkel2020} & $759_{-161}^{+252}$ & 266 & 235 & 1.4 & N \\
Theia 1098 & $589_{-172}^{+243}$ & \citet{Kounkel2020} & $1562_{-295}^{+445}$ & 312 & 144 & 0.2 & Y \\
Theia 1358 & $1122_{-363}^{+538}$ & \citet{Kounkel2020} & $1772_{-393}^{+609}$ & 371 & 173 & -0.5 & N \\
Theia 1532 & $1445_{-398}^{+550}$ & \citet{Kounkel2020} & $1785_{-359}^{+541}$ & 403 & 233 & 0.0 & Y \\
MELANGE-1 & $250_{-70}^{+50}$ & \citet{THYMEV}  & $200_{-68}^{+168}$ & 111 & 35 & 1.9 & ... \\
MELANGE-2 & $700$ & \citet{THYMEVII} & $816_{-269}^{+584}$ & 119 & 32 & 0.0 & ... \\
MELANGE-3 & $105_{-10}^{+10}$ & \citet{Barber2022} & $296_{-50}^{+69}$ & 326 & 321 & 33.6 & ... \\
MELANGE-4 & $27_{-3}^{+3}$ & \citet{Wood2022} & $26_{-5}^{+8}$ & 94 & 101 & 59.4 & ... \\
MELANGE-6 & $150_{-25}^{+25}$ & \citet{Vowell2023} & $117_{-25}^{+39}$ & 153 & 104 & 16.0 & ...\\ 
\enddata
\tablenotetext{a}{Ages for the Theia groups are converted from dex values}
\tablenotetext{b}{Median distance of members rather than reported distance.}
\tablenotetext{c}{$N_{\rm{stars}}$ denotes the number of stars used in our analysis (after applying all cuts). The full membership list is always larger.}
\end{deluxetable*}

\bibliography{Gaia_var_mann}{}
\bibliographystyle{aasjournal}

\clearpage

\end{document}